\documentclass[%
aps,
prx,
reprint,
superscriptaddress,
nofootinbib,
amsmath,
amssymb
]{revtex4-2}
\usepackage{
    physics,
    graphicx,
    multirow,
    tabularx,
    mathtools, 
    placeins, 
    nicefrac, 
    hyperref, 
    threeparttable, 
    siunitx,
    enumitem,
    makecell
}

\newcommand{\rowhead}[2]{\parbox[t][5ex][t]{#1}{\footnotesize #2}}

\usepackage{tikz}
\usetikzlibrary{quantikz2}

\usepackage{amsthm}

\usepackage{dsfont}

\usepackage[capitalize]{cleveref}
\crefname{section}{Sec.}{Secs.}
\Crefname{section}{Sec.}{Secs.}
\crefname{appendix}{Appendix}{Appendices}
\Crefname{appendix}{Appendix}{Appendices}

\usepackage[dvipsnames]{xcolor}
\hypersetup{
    colorlinks,
    linkcolor={blue!50!black},
    citecolor={blue!50!black},
    urlcolor={blue!80!black}
}

\usepackage[caption=false]{subfig} 
\DeclarePairedDelimiterX\pbraket[2]{\langle\!\langle}{\rangle\!\rangle}{#1 \delimsize\vert #2}

\begin{document}

\title{Correcting quantum errors using a classical code\\and one additional qubit}

\newcommand{\qmaddress}{\affiliation{Quantum Motion, 9 Sterling Way, London N7 9HJ, United Kingdom}}
\newcommand{\oxddress}{\affiliation{Department of Engineering Science, University of Oxford, Parks Road, Oxford OX1 3PJ, United Kingdom}}

\author{Tenzan Araki}
\email{tenzan.araki@physics.ox.ac.uk}
\affiliation{Department of Physics, Clarendon Laboratory, University of Oxford, Parks Road, Oxford OX1 3PU, United Kingdom}
\affiliation{Mathematical Institute, University of Oxford, Woodstock Road, Oxford OX2 6GG, United Kingdom}
\author{Joseph F. Goodwin}
\affiliation{Department of Physics, Clarendon Laboratory, University of Oxford, Parks Road, Oxford OX1 3PU, United Kingdom}
\author{Zhenyu Cai}
\email{cai.zhenyu.physics@gmail.com}
\oxddress
\qmaddress

\begin{abstract}
Classical error-correcting codes are powerful but incompatible with quantum noise, which includes both bit-flips and phase-flips. We introduce Hadamard-based Virtual Error Correction (H-VEC), a protocol that empowers any classical bit-flip code to correct Pauli noise with the addition of only a single control qubit and two layers of controlled-Hadamard gates. Through classical post-processing, H-VEC virtually filters the error channel, projecting the noise into pure $Y$-type errors that are subsequently corrected using the classical code's native decoding algorithm. We demonstrate this by applying H-VEC to the classical repetition code. Under a code-capacity noise model, the resulting protocol not only provides full quantum protection but also achieves an exponentially stronger error suppression (in distance) than the original classical code. The improvements over the surface code are even more pronounced, while using far fewer qubits, simpler checks, and straightforward decoding. Considering circuit-level noise, we present a fault-tolerant protocol in which H-VEC can quadratically reduce the qubits needed for long-range surface code lattice surgery. There are some limitations to the technique, most notably that H-VEC introduces a sampling overhead due to its post-processing nature. Nonetheless, it represents a fundamentally novel hybrid quantum error correction and mitigation framework that redefines the trade-offs between physical hardware requirements and classical processing for error suppression.
\end{abstract}

\maketitle

\section{Introduction \label{sec: intro}}
Classical error correction (CEC) has underpinned reliable digital computation and communication for decades. It is cheap and robust at almost all practical scales thanks to a mature ecosystem of efficient codes and decoders. However, its machinery cannot be directly ported to quantum error correction (QEC) without substantial conceptual and practical overhead. While classical channels involve only bit‑flip errors, quantum devices face both bit‑ and phase‑flip errors and thus require inference of both error location and type. Such a fundamental difference prevents direct implementation of CEC in the quantum setting, leading to challenges such as the compatibility between error checks and complicated decoding problems due to error degeneracy. 

The CSS construction~\cite{calderbank1996good,steane1996multiple} partially reconciles classical structure with quantum requirements, but at the cost of extra qubits, circuitry, connectivity, and decoding overheads, widening the resource requirement gap between CEC and fully fledged QEC. Indeed, vast efforts have been poured into finding more efficient constructions of quantum codes that can rival the desirable properties found in CEC~\cite{breuckmannQuantumLowdensityParitycheck2021}, e.g., the hypergraph‑product construction entered the quantum LDPC paradigm with non‑vanishing rate and distance scaling with $\sqrt{n}$~\cite{tillichQuantumLDPCCodes2014}; hyperbolic surface codes offer constant rate, albeit with logarithmic distance~\cite{breuckmannConstructionsNoiseThreshold2016}; and the lifted‑product families can achieve a range of trade-offs between rates and distance~\cite{panteleevQuantumLDPCCodes2022}.  All of these developments culminated in the discovery of asymptotically good quantum LDPC codes~\cite{panteleevAsymptoticallyGoodQuantum2022,dinurLocallyTestableCodes2022}, achieving both linear rate and distance in the asymptotic limit, just as many of its classical counterparts. However,  while these codes bring potentially significant reductions in the qubit overhead, they pose additional challenges in terms of decoding complexity, stabiliser measurement circuits, and connectivity requirements. There is still a long way ahead for constructing a practical quantum error correction protocol that can rival the simplicity, efficiency, and practical performance of classical codes.

It is thus desirable to develop new ways to leverage the vast wealth of existing CEC protocols for more resource-efficient ways to correct quantum noise. Most recently, Ref.~\cite{liu2025virtual} introduced SWAP-based \emph{virtual error correction (VEC)}, which allows the direct use of classical codes for correcting quantum noise with the help of post-processing and controlled-SWAP gates. In that protocol, two classical codes are required to protect against bit-flip and phase-flip errors, respectively, which is very close to resolving a seemingly impossible proposition:

\ \\
\emph{Is it possible to take a single classical code, keeping its native qubit count and distance scaling, and empower it to correct the full spectrum of quantum noise?}
\ \\

In this work, we show that this is indeed possible using the \emph{Hadamard‑based VEC (H-VEC)} protocol, in which we only need to add a single control qubit and two layers of controlled‑Hadamard (C‑H) gates to a classically encoded register to protect against quantum noise (as shown in \cref{fig: vqc}). Our protocol not only neutralises phase errors beyond the reach of the input classical code, but also yields stronger bit-flip error suppression than the input code. Under local depolarising noise, applying H-VEC to the bit-flip repetition code can achieve exponentially (in distance) smaller logical bit-flip error compared to the constituent bit-flip code, and achieve an even larger factor of improvement when compared to surface codes, despite using far fewer qubits. 

As a protocol based on post-processing, H-VEC provides a native integration between quantum error mitigation (QEM)~\cite{cai2023quantum} and QEC. In many cases, our protocol can be thought of as an efficient and general way to effectively filter out the $X$ and $Z$ error components, leaving only $Y$ errors, which are correctable by the classical code. Such post-selection naturally comes with a sampling overhead and is not meant to be scaled indefinitely. Instead, it should be viewed as a protocol that enables us to extract more power out of whatever hardware we have, either via direct implementation or by combining it with other QEC schemes. For example, as we will demonstrate, the latter application provides a practical fault-tolerant scheme for long-range lattice surgery that operates reliably under circuit-level noise while achieving quadratically improved qubit costs. H-VEC presents a new way to trade sampling overhead for a drastic reduction in the requirements on qubit numbers, error check circuits, connectivity and decoding, which can be useful in many contexts. While we derive the scheme assuming an application to expectation value problems, we expect the results to hold in more general settings such as for sampling problems~\cite{liu2025quantum}.

We introduce H-VEC and derive its mechanism in \cref{sec: vqc}. This is followed by \cref{sec:performance}, in which we analyse the performance of H-VEC under local depolarising noise and compare it with the repetition and surface codes, with results summarised in \cref{fig: threshold-cc} and \cref{tab:comparison}. We describe a practical application of H-VEC in \cref{sec: interact}, demonstrating how H-VEC interacts with QEC to enable fault-tolerant long-range lattice surgery under a circuit-level noise model. \cref{sec: general} presents a general VEC framework that subsumes H-VEC and the SWAP-based variant (SWAP-VEC). We conclude in \cref{sec:discussion} with discussions and the potential outlook. Detailed practical considerations of H-VEC, such as its implementation in biased-noise systems, under connectivity constraints, and for protecting resource states, are deferred to the Appendices.

\section{Correcting quantum errors using classical codes}\label{sec: vqc}

\begin{figure}
\includegraphics[width=0.48\textwidth]{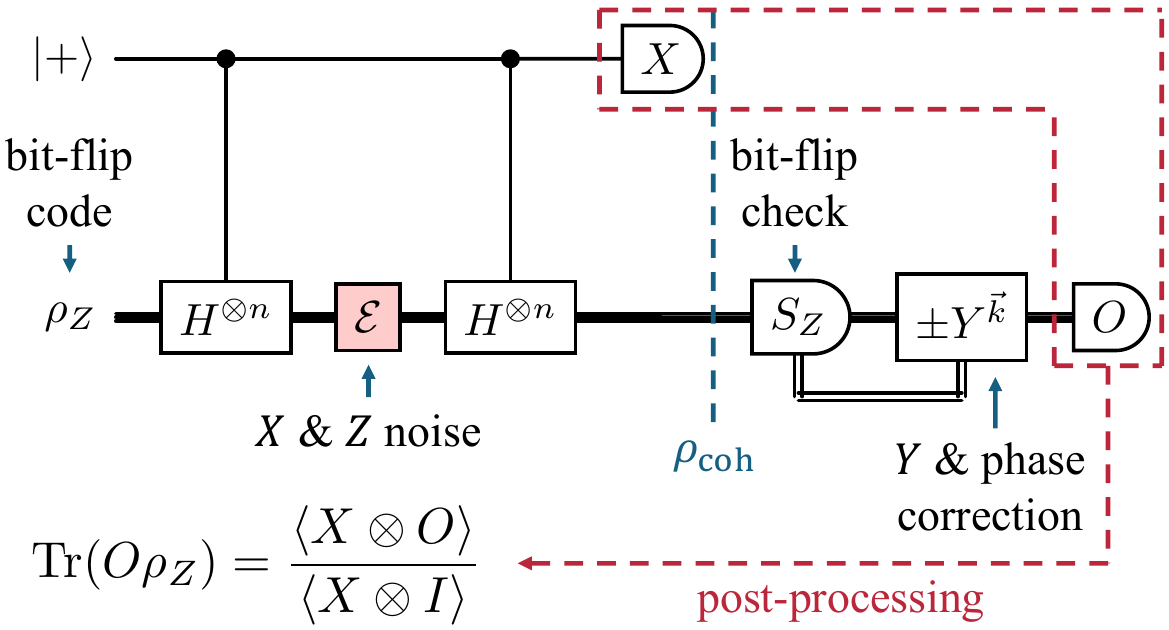}
\caption{A diagram showing the H-VEC scheme presented in this work. The objective is to obtain an observable expectation value Tr$(O\rho_Z)$, where $\rho_Z$ is a bit-flip classical code state encoded in an $n$-physical-qubit logical register that suffers from an error channel $\mathcal{E}$, which comprises of both bit-flip and phase-flip errors. By initialising a control qubit in $\ket{+}=(1/\sqrt{2})(\ket{0}+\ket{1})$ and sandwiching the error channel between two layers of C-H gates, the noise on the logical register, after the $X$ measurement on the control qubit and the bit-flip checks $S_{Z}$, is effectively projected into one that consists purely of $Y$-type errors. The projected errors can then be corrected using $Y$ gates via the measured syndrome $\vec{k}$, along with phase adjustments performed in classical software. The desired error-corrected expectation value can then be retrieved via the ratio of the two expectation values shown.}\label{fig: vqc}
\end{figure}

In classical computation, data bits only suffer from bit-flip errors. Let us denote bit-flip errors as $X^{\vec{x}} = \prod_{i=1}^{n} X_i^{x_i}$, where $n$ is the number of bits, $X$ is the bit-flip operator and $\vec{x}$ is a $n$-bit binary vector that specifies the error locations. Given a CEC code defined by the code space projector $\Pi_Z$ (with a subscript $Z$ since it is defined by $Z$ stabilisers), we denote the set of errors correctable by this code as $\mathbb{E}_{\mathrm{cor}}$, and they follow the Knill-Laflamme condition~\cite{knill1997theory} for classical codes:
\begin{align}\label{eqn:kl_classical}
    \Pi_{Z} X^{\vec{x}}X^{\vec{v}} \Pi_{Z} = \delta_{\vec{x}, \vec{v}} \Pi_{Z} \quad \forall \vec{x}, \vec{v} \in \mathbb{E}_{\mathrm{cor}}.
\end{align}

In quantum computation, qubits also suffer from phase-flip errors, denoted by $Z^{\vec{z}}$. A classical code as described above consists only of $Z$-type stabilisers, and is thus unable to correct any phase-flip errors as they commute with all stabilisers (error checks). One may add $X$ stabilisers using the CSS formalism~\cite{calderbank1996good,steane1996multiple}, but as mentioned in \cref{sec: intro}, such codes are non-trivial to construct and come with additional overheads.

Here we introduce the \emph{Hadamard‑based VEC (H-VEC)} scheme, which uses an existing classical code to correct for both bit-flip and phase-flip errors. The scheme involves only one additional qubit by leveraging the quantum circuit shown in \cref{fig: vqc}. The input is some classical bit-flip code state $\rho_Z$ in the code space defined by $\Pi_{Z}$ and the control qubit is prepared in $\ket{+}=(1/\sqrt{2})(\ket{0}+\ket{1})$. We then apply two layers of controlled-Hadamard (C-H) gates surrounding the noise channel that we wish to correct. This noise channel we consider takes the form
\begin{align}\label{eq: pauli}
    \mathcal{E}[\;\cdot\;] = \sum_{\vec{x}, \vec{{z}} \in \mathbb{E}_{\mathrm{cor}}}p_{\vec{x}, \vec{z}} X^{\vec{x}} Z^{\vec{z}} \,\cdot\, Z^{\vec{z}} X^{\vec{x}},
\end{align}
which contains both $X$ and $Z$ noise and thus is beyond the reach of classical codes. Our arguments can also be extended to the most general Pauli channel with error locations $\vec{x}$ and $\vec{{z}}$ not restricted to within the correctable set $\mathbb{E}_{\mathrm{cor}}$. Such a general channel beyond $\mathbb{E}_{\mathrm{cor}}$ is only necessary for calculating the exact logical error rates as outlined in \cref{sec:uncorrect_error}. In this section, for the purpose of demonstrating our protocol's ability to correct for phase-flip errors using an input bit-flip code state, the error model in \cref{eq: pauli} is sufficient and can provide much simplified analysis.

As we run the quantum circuit shown in \cref{fig: vqc}, we multiply a factor of $\pm 1$ obtained from measuring the control qubit with the output of the main register. The post-processed effective output ``state'' at the vertical dashed line can be expressed as (see \cref{sec:derivation})
\begin{align}\label{eq: virtual state}
    \rho_{\mathrm{coh}} \propto \sum_{\vec{x}, \vec{{z}} \in \mathbb{E}_{\mathrm{cor}}}p_{\vec{x}, \vec{z}} X^{\vec{x}} Z^{\vec{z}} \rho_Z  X^{\vec{z}} Z^{\vec{x}}  + c.c.
\end{align}
This is referred to as a \textit{virtual} state, since it is obtained via post-processing and is not necessarily positive semi-definite. Here, the Kraus operator on one side has been transformed by transversal Hadamard operations. Note that this takes a similar form as the cross term (coherent off-diagonal component) in the presence of coherent noise of form $\propto X^{\vec{x}} Z^{\vec{z}} + X^{\vec{z}} Z^{\vec{x}}$.

Now, if we perform a stabiliser measurement for the bit-flip code on this virtual state as shown in \cref{fig: vqc}, the virtual state would be projected into the syndrome subspace $X^{\vec{k}} \Pi_Z X^{\vec{k}}$, where $\vec{k} \in \mathbb{E}_{\mathrm{cor}}$ is the label for the resultant syndrome. Applying this projector to the virtual state in \cref{eq: virtual state} and using the Knill-Laflamme condition in \cref{eqn:kl_classical}, the resulting effective output state is given by
\begin{align}\label{eq: projected state}
    \rho_{\vec{k}} &\propto \left(X^{\vec{k}}\Pi_{Z}X^{\vec{k}}\right) \rho_{\mathrm{coh}} \left(X^{\vec{k}}\Pi_{Z}X^{\vec{k}}\right) \nonumber\\
    & = \sum_{\vec{x}, \vec{z} \in \mathbb{E}_{\mathrm{cor}}}  \delta_{ \vec{x}, \vec{k}} \delta_{ \vec{z}, \vec{k}} (-1)^{\vec{x}\cdot\vec{z}}p_{\vec{x}, \vec{z}} X^{\vec{k}} Z^{\vec{z}} \rho_Z  Z^{\vec{x}}X^{\vec{k}} \nonumber\\
    &= (-1)^{|\vec{k}|} p_{\vec{k}, \vec{k}} Y^{\vec{k}} \rho_Z  Y^{\vec{k}},
\end{align}
where $\vec{x}\cdot\vec{z}$ denotes a binary dot product between $\vec{x}$ and $\vec{z}$ and $|\vec{k}| = \vec{k} \cdot \vec{k}$ is the Hamming weight of $\vec{k}$. Hence, the bit-flip stabiliser measurement ($Z$ checks) collapses the errors into $Y^{\vec{k}}$, which can be corrected by simply applying $Y^{\vec{k}}$ and a $(-1)^{|\vec{k}|}$ phase to the output state.

Following the procedure above, for some observable of interest $O$, one can obtain its noiseless expectation value on the input code state $\rho_Z$ using the circuit in \cref{fig: vqc} by measuring (see \cref{sec:derivation})
\begin{align}\label{eq: final}
    \Tr(O \rho_Z) = \frac{\expval{X \otimes O}}{\expval{X \otimes I}}.
\end{align}
The expectation values on the right-hand side are obtained by performing the respective measurements in \cref{fig: vqc}. The denominator is a normalisation factor given by (see \cref{sec:derivation})
\begin{align*}
    \expval{X \otimes I} = \sum_{\vec{k}\in \mathbb{E}_{\mathrm{cor}}} p_{\vec{k}, \vec{k}} = P_Y,
\end{align*}
where we use $P_\sigma$ to denote the total probability of purely $\sigma$-type errors occurring for $\sigma \in \{X,Y,Z\}$ (Note that identity, i.e. no errors, is included in any of these $\sigma$-type errors as the zero-weight error). Following similar arguments in Refs.~\cite{koczor2021exponential,huggins2021virtual,caiPracticalFrameworkQuantum2021} using Hoeffding’s inequality, the sampling overhead (i.e., factor of increase in the number of circuit runs) of our post-processing scheme is given by
\begin{align*}
    C_Y \approx P_Y^{-2}.
\end{align*}

This shows that, if we have a noise model that is heavily biased towards $Y$, then we would have very small sampling overhead. The error suppression power would correspondingly also move closer to that of the classical input code, which is already very strong, and we still get to maintain the ability to correct (or, more precisely, filter out) the $X$ and $Z$ noise. Hence, our protocol can be a very effective way to go one step beyond classical codes in the biased noise regime, among more general application scenarios. It should be emphasised that we focus on $Y$ noise here purely for pedagogical simplicity in the introduction of our techniques, while bias towards dephasing ($Z$) is more common. Fortunately, the above result generalises straightforwardly to different Pauli bases $\sigma$ by adjusting the input classical code and the C-H gates (see \cref{sec: bias}). This yields a sampling overhead of $P_\sigma^{-2}$, enabling us to select the $\sigma$-variant in the presence of noise biased toward $\sigma$-type errors.

\section{Performance}\label{sec:performance}
To provide a concrete illustration of the H-VEC protocol, let us consider representative example codes and assess their performances under the code-capacity error model, in which stabiliser checks are assumed perfect. Specifically, we will use the classical bit-flip repetition code as input for H-VEC, constructing what we call a \textit{virtual quantum repetition code}. We will compare it against the repetition code and the surface code, where the latter is the direct quantum analogue of the former, since it is the hypergraph product of two repetition codes~\cite{krishnaFaulttolerantGatesHypergraph2021}. Following a detailed comparison of their performance, we will discuss some possible limitations of our protocol. The results, including resource requirements, are summarised in \cref{tab:comparison}.

\subsection{Code Performance under Depolarising Noise}\label{sec:example}
Consider distance-$d$ codes (with $d$ being an \emph{odd number} for simpler expressions) and a noise model whereby each physical qubit experiences a single-qubit depolarising noise of the form $\mathcal{D}_p[\,\cdot\,] = (1-p)I\cdot I + \frac{p}{3}\sum_{G\in \{X,Y,Z\}} G\cdot G$, where $p$ is the physical error rate. In the following, we will study the performance of the virtual quantum repetition code, along with the repetition code and the surface code. We will use the logical $\ket{+}_{\mathrm{L}}$ ($\ket{0}_\mathrm{L}$) state as the input for detecting logical $Z$ ($X$) errors, which will also include the error rate of logical $Y$ errors since they have a $Z$ ($X$) component. For all of our expressions of logical error rate here, we will take the limit of small $p$ and keep only the leading-order term to observe their scaling behaviour. More detailed analysis and derivation of the error rates can be found in \cref{sec:example_derivation}.

\paragraph{Repetition code.} We need $d$ physical qubits to encode one logical qubit. It can correct all $X$ errors up to weight $(d-1)/2$ (i.e. $\mathbb{E}_{\mathrm{cor}}$ contains all the bit strings that have Hamming weights up to $(d-1)/2$). The leading-order probability of logical $X$ errors is given by
\begin{align}\label{eqn:p_l_rep_x}
    p_{\textrm{L,rep}, X}
    &\approx \binom{d}{(d+1)/2} \left(1-\frac{2p}{3}\right)^{\frac{d-1}{2}} \left(\frac{2p}{3}\right)^{\frac{d+1}{2}}.
\end{align}
It cannot correct any $Z$ error components (which is also contained in $Y$ errors). The probability of logical $Z$ errors is given by
\begin{align}\label{eqn:p_l_rep_z}
    p_{\textrm{L,rep},Z} & = 1 - \left(1-\frac{2p}{3}\right)^d \approx \frac{2dp}{3}
\end{align}

\paragraph{Surface code. } We need $d^2 + (d-1)^2$ physical qubits to encode one logical qubit for the unrotated surface code. It has the same error correcting power for both $X$ and $Z$ errors and can correct all errors up to weight $(d-1)/2$. Its logical error probability, keeping only the leading-order weight-$(d+1)/2$ term, can be approximated by (see \cref{sec:example_derivation})
\begin{align}\label{eqn:p_l_sur}
    p_{\textrm{L,sur}} &\approx \left(5d-4\right) \binom{d}{(d+1)/2} \left(1-\frac{2p}{3}\right)^{\frac{d-1}{2}} \left(\frac{2p}{3}\right)^{\frac{d+1}{2}} \nonumber\\
    & \approx \left(5d-4\right) p_{\textrm{L,rep}, X}
\end{align}

\paragraph{Virtual quantum repetition code.} We need $d$ physical qubits to encode one logical qubit of the repetition code and an additional qubit as the control qubit. It has equal error correcting power for both $X$ and $Z$ errors. As derived in \cref{sec:example_derivation}, the logical error rate is approximately
\begin{align}\label{eqn:p_l_vec}
    p_{\textrm{L,vec}} \approx \left(1-\frac{2p}{3}\right)^{-d} \binom{d}{(d+1)/2}  (1-p)^{\frac{d-1}{2}} \left(\frac{p}{3}\right)^{\frac{d+1}{2}}.
\end{align}
As mentioned at the end of \cref{sec: vqc}, the sampling overhead is related to the probability of pure $Y$ errors, which is given by
\begin{align*}
    P_{Y} &= \left(1 - \frac{2p}{3}\right)^{d}.
\end{align*}
Hence, the sampling overhead factor is approximately
\begin{align}\label{eqn:vec_example_overhead}
    C_{Y} &\approx P_{Y}^{-2} \approx \left(1 - \frac{2p}{3}\right)^{-2d}.
\end{align}

\subsection{Stronger Error Suppression using H-VEC}
\begin{figure*}
    \centering
    \includegraphics[width=0.9\textwidth]{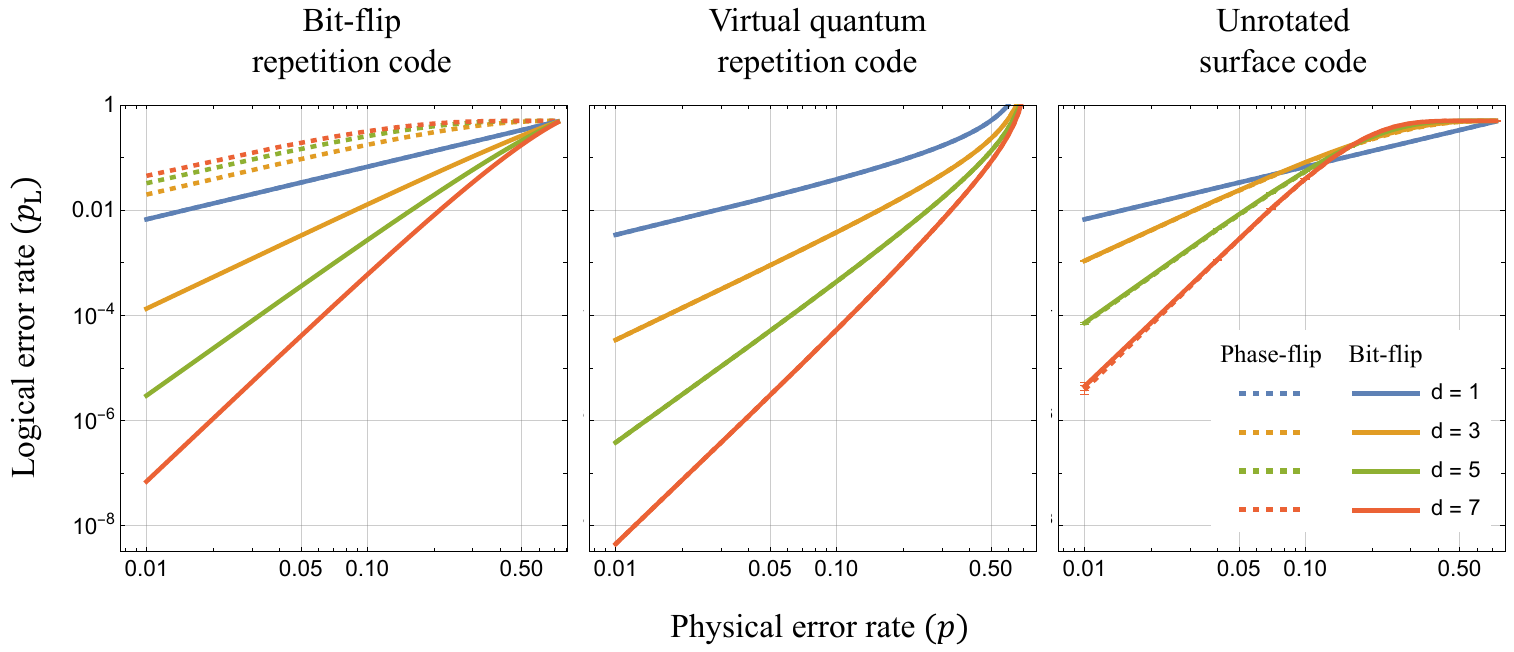}
    \caption{Logical error rate $p_{\mathrm{L}}$ as a function of physical error rate $p$ simulated under the code-capacity error model with local depolarising errors, comparing the bit-flip repetition code (left), virtual quantum repetition code (middle), and unrotated surface code (right) for code distances $d \in \{1,3,5,7\}$. In contrast to the bit-flip repetition code, the virtual quantum repetition code is able to suppress both bit-flip and phase-flip errors, and it does so more effectively than the surface code. See \cref{app: numerics} for simulation details.}
    \label{fig: threshold-cc}
\end{figure*}
In the above results, it is not surprising that the H-VEC scheme can suppress phase-flip noise much better than the classical bit-flip code since the bit-flip code is not protected against phase errors at all. However, when we compare the bit-flip error protection in \cref{eqn:p_l_rep_x,eqn:p_l_vec}, we obtain
\begin{align}
    \frac{p_{\textrm{L,rep}, X}} {p_{\textrm{L,vec}}} = 2^{\frac{d+1}{2}} \left(1+\order{p}\right),
\end{align}
i.e., the bit-flip error rate of the virtual repetition code is exponentially smaller (in $d$) than the bit-flip repetition code of the same distance. Such a strong error suppression power is achieved for both $X$ and $Z$ logical errors without resorting to full quantum error correction. In fact, the improvement of its error suppression power over the surface code is even larger:
\begin{align}
    \frac{p_{\textrm{L,sur}}} {p_{\textrm{L,vec}}} = (5d-4)2^{\frac{d+1}{2}} \left(1+\order{p}\right).
\end{align}
This is because there are more ways for a surface code to fail due to the increased number of qubits. 

In \cref{fig: threshold-cc}, we show logical error rates of the three different protocols obtained from numerical simulations (detailed in \cref{app: numerics}). Indeed, we observe the expected stronger error suppression achieved by the virtual quantum repetition code over both the repetition code and the surface code in both $X$ and $Z$ errors. These numerical results also closely align with the analytical expressions derived in \cref{sec:example} as shown in \cref{fig: threshold-theory}.

The most important reason for the stronger error suppression of the H-VEC scheme is because, as discussed in \cref{sec: vqc}, we are effectively removing most of the errors that are not pure $Y$ errors in the process, such that to a very good approximation, only pure $Y$ errors can contribute to the logical errors. On the other hand, both $Y$ and $X$ errors can contribute to logical bit flip for both the repetition code and surface code. This resulted in the $\left(p/3\right)^{\frac{d+1}{2}}$ factor in \cref{eqn:p_l_vec} for the virtual quantum repetition code and the $\left(2p/3\right)^{\frac{d+1}{2}}$ factor in \cref{eqn:p_l_rep_x,eqn:p_l_sur}, improving the error suppression power by a factor of $2^{\frac{d+1}{2}}$. Hence, we can also view our protocol as a way to (effectively) post-select our circuit runs to filter out any instance with errors that are not purely of $Y$-type. This enables correction using only classical codes and also achieves exponentially stronger error suppression with increasing distance. 

\begin{table}[tbhp!]
    \centering
    \renewcommand{\arraystretch}{2}
    \setlength{\tabcolsep}{0.5em}
    \begin{tabular}{c | ccc}\toprule
            &Repetition & Virtual & Surface\\\hline
            Qubits& $d$ & $d+1$ & $d^2 + (d-1)^2$\\
            \rowhead{14ex}{Relative Logical Error}&$1$ (only $X$)&$2^{-\frac{d+1}{2}}$&$5d-4$\\
            \rowhead{10ex}{Sampling Costs}&$1$&$\left(1 - \frac{2p}{3}\right)^{-2d}$&$1$\\
            \botrule
    \end{tabular}
    \caption{Comparison between distance-$d$ bit-flip repetition code, virtual quantum repetition code and unrotated surface code. The relative logical error rate here is in the unit of $p_{\textrm{L,rep}, X}$ in \cref{eqn:p_l_rep_x}, which is the bit-flip logical error rate achieved by the bit-flip repetition code. The bit-flip repetition code cannot correct any phase errors, so the logical error rate here is only for bit-flip errors. The logical error rate for surface code here is not the exact figure but an analytical approximation.}\label{tab:comparison}
\end{table}

\subsection{Cost of Implementation \label{subsec: limits}}
The effective post-selection for pure $Y$ errors in H-VEC naturally comes with a sampling overhead. If such post-selection can indeed be performed directly, then the overhead will be $\left(1 - 2p/3\right)^{-d}$, which is the inverse of the probability of getting pure $Y$ errors. However, since we are achieving the same effect through post-processing instead of direct post-selection, our sampling overhead is quadratically higher at $\left(1 - 2p/3\right)^{-2d}$. As we can see in \cref{fig: sampling-cost}, the sampling overhead is manageable for sufficiently small physical error rates, where the latter is already the regime of practical interest for enabling QEC.

Compared to a direct application of the repetition code, our protocol requires one additional control qubit and $2d$ C-H gates, whose noise we have not taken into account of in the above analysis. First of all, some of this noise can just be merged into the noise channel that we can mitigate in between the C-H layers. For the noise on the control qubit, Ref.~\cite{linErrormitigatedQuantumMetrology2025} has shown that under many common noise types, including depolarising, dephasing and amplitude damping, the effect of control qubit noise on the numerator and denominator in \cref{eq: final} will cancel each other and thus will not affect our protocol (see \cref{app: robustness} for details). For the remaining noise, one way to combat them is to apply additional error mitigation techniques, such as probabilistic error cancellation (PEC), which has been shown to play nicely with such a virtual error correction scheme~\cite{liu2025virtual}. While this will require good noise characterisation of the C-H gates, it should not pose significant challenges since there is only one gate type to characterise. Moreover, the additional sampling overhead will be small, since there is only a linear number of C-H gates for which we mitigate errors.

The additional gates introduced by H-VEC also imply that a high degree of connectivity between the control qubit and the physical qubits in the logical register would be desired, and the sequential implementation of each C-H gate would translate to an increased circuit depth of $\mathcal{O}(n)$. While it is possible to apply the C-H gates in parallel by fanning-out~\cite{haner2021distributed} the control qubit into $n$ control qubits, this would impose the additional overhead of preparing them in an $n$-qubit Greenberger–Horne–Zeilinger (GHZ) state. In \cref{app: transversal}, we show that it is in fact possible to replace the single control qubit with $n$ \textit{unentangled} ones to implement each C-H layer in a single parallelised step as illustrated in \cref{fig: CH_decomp} without altering the performance and sampling overhead of H-VEC. Importantly, there is no need to create a GHZ state, and thus no direct interactions between the control qubits are required. The resulting circuit is flexible in terms of connectivity requirements, allowing H-VEC to be compatible with devices where connectivity between qubits comes at a non-negligible cost~\cite{itogawa2025efficient,pino2021demonstration,nickerson2013topological,nickerson2014freely}.

\begin{figure*}
\centering
    \includegraphics[width=0.98\textwidth]{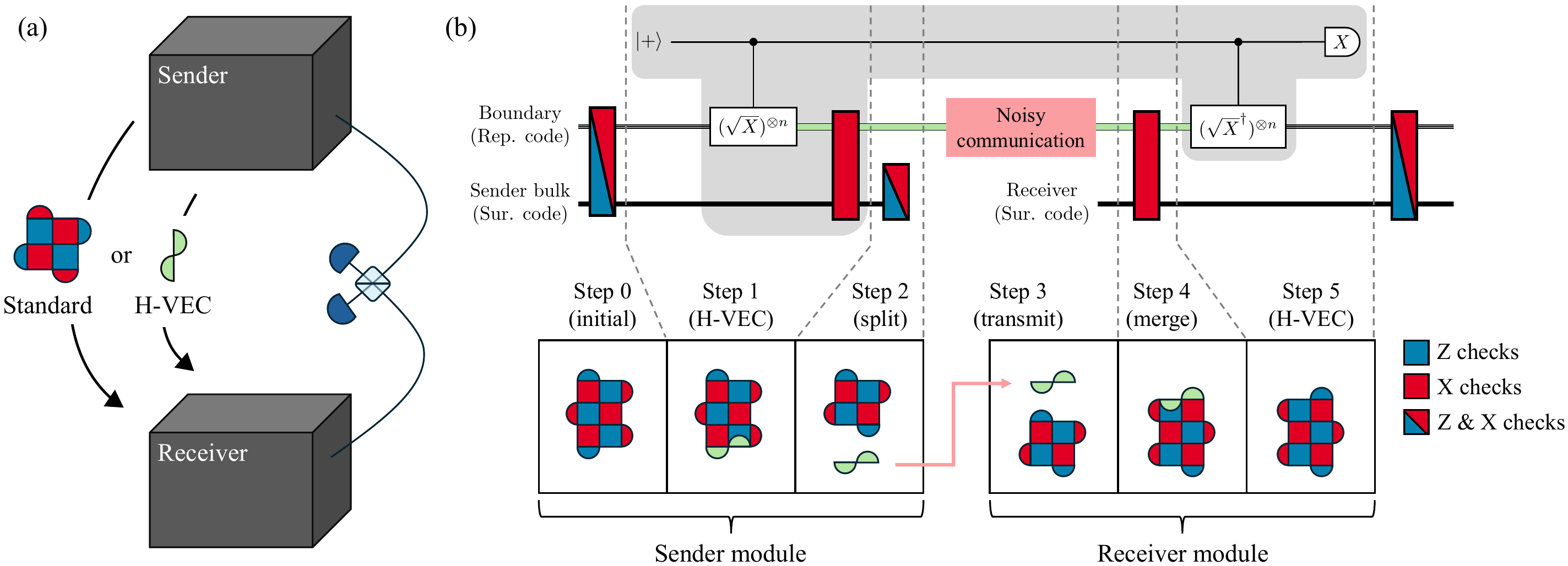}
    \caption{\textbf{Protocols for long-range lattice surgery between surface codes.} \textbf{(a)} Problem setup with two modules connected by shared Bell pairs that enable long-range lattice surgery through transmitting the entire code patch in the standard approach, or transmitting only the boundary qubits in H-VEC. \textbf{(b)} Detailed circuit and procedure for carrying out long-range lattice surgery by sending only $\mathcal{O}(d)$ (rather than $\mathcal{O}(d^2)$) physical qubits without H-VEC (non-fault-tolerant, without the components in the shaded area) and with H-VEC (with the shaded components). The quantum circuit is shown on the top and the code patches immediately following each step on the bottom. Qubits coloured in green are protected by H-VEC if the shaded components are applied. Step 1: Apply the first C-$\sqrt{X}$ layer to a boundary of the sender code patch to enter the protected region of H-VEC, and apply $X$ checks to suppress any phase errors introduced by the former gate layer. Step 2: Split the boundary from the bulk of the sender code patch. Step 3: Transmit the detached boundary through the noisy communication channel. Step 4: Merge the boundary with the receiver code patch. Step 5: Apply the second C-$\sqrt{X}$ layer to exit the protected region of H-VEC. The robustness of the protocol against the additional noise introduced by the C-$\sqrt{X}$ gates and the control qubit is discussed in \cref{sec: ft_scheme}.
    }\label{fig: surgery}
\end{figure*}

\section{Fault-tolerant Application\label{sec: interact}}

So far, we have limited ourselves to the code capacity noise model. In practice, since the stabiliser checks lie outside the two C-H layers, they are not fully protected against quantum noise. One way to deal with this is to incorporate additional QEM. The associated sampling cost will be reasonable due to the fact that we are performing classical error checks, which can be much simpler and thus much less noisy compared to their quantum counterparts. A much more practical approach is to replace critical steps in existing fault-tolerant protocols with H-VEC. This may offer protection against general noise sources beyond those natively suppressed by H-VEC, while still enjoying the resource-efficiency of H-VEC. 

In this section, we will see how the virtual quantum repetition code, introduced in \cref{sec:performance}, can enable fault-tolerant long-range lattice surgery \cite{horsman2012surface,fowler2019low,litinski2019game,vuillot2019code} for the surface code under a \emph{circuit noise model} (summarised in \cref{fig: surgery}), leading to quadratically improved qubit costs. The method can potentially be extended to facilitate interaction between more general quantum codes using its constituent classical code empowered by H-VEC, which can be an interesting future direction. Our protocol involves performing further QEC processes after our QEM circuit. This makes use of the fact that QEC can be directly applied on top of physical qubits that have been error-mitigated using linear QEM methods like H-VEC, with the effective physical noise model being the resultant error-mitigated noise model from the given QEM techniques~\cite{jeon2026quantum}.

\subsection{Background}\label{sec: surgery-background}
Lattice surgery was initially proposed as a means to perform logical operations in surface code architectures constrained to nearest-neighbor connectivities, and its implementation involves merging two code patches through ancilla-assisted syndrome measurements to obtain their logical parity \cite{vuillot2019code}. When the two code patches are hosted in distinct modules of a modular quantum device, care must be taken to retain fault-tolerance while consuming as few remote Bell pairs as possible, as the latter are typically the bottleneck of such architecture both in terms of fidelity and rate. If we use these Bell pairs to directly facilitate the long-range gates used to perform lattice surgery between the two surface code patches, then we will need $\mathcal{O}(d)$ rounds of long-range syndrome extraction with $\mathcal{O}(d)$ Bell pairs consumed in each round~\cite{shalby2025optimized, jacinto2026network, haug2025lattice}, leading to a consumption of $\mathcal{O}(d^2)$ Bell pairs in total. Alternatively, one can transmit one of the two code patches from a \textit{sender} module to a \textit{receiver} module through quantum teleportation as the standard approach depicted in \cref{fig: surgery}~(a). This would allow for an arbitrary number of local syndrome measurement rounds once the transmission is complete, but it still requires $\mathcal{O}(d^2)$ Bell pairs for the transmission, and also $\mathcal{O}(d^2)$ additional qubits in the receiver module to host the entire transmitted code patch.

Since the two code patches only interact at their boundaries, one may ask whether it is sufficient to only transmit the $\mathcal{O}(d)$ data qubits that correspond to the relevant boundary of one of the code patches, rather than the entire $\mathcal{O}(d^2)$ data qubits. Without loss of generality, let us consider transmitting the $Z$ boundary (the boundary stabilised by $Z$ operators) of a rotated surface code to meet the $Z$ boundary of another. As shown in \cref{fig: surgery}~(b), ignoring the shaded components in the circuit corresponding to Steps 1 and 5, this would begin with the boundary splitting off from the sender code patch via new weight-2 $Z$ checks in the bulk, followed by transmitting only the boundary to the receiver module before merging it with the receiver code patch through new $X$ checks.

Unfortunately, a na\"{i}ve implementation of this approach as described above is not fault-tolerant since the boundary to be transmitted through the noisy channel is a bit-flip repetition code of distance $d$ that cannot suppress phase noise. Hence, any single-qubit $Z$ error during splitting, transmitting and merging on these boundary qubits will lead to a logical error. However, as we will show below, we can use H-VEC to protect these boundary qubits in all of these steps against both bit-flip and phase-flip errors, giving rise to a fault-tolerant protocol.

\subsection{Fault-tolerant Scheme Enabled by H-VEC}\label{sec: ft_scheme}
Let us consider the $X$-basis version of H-VEC, where the C-H gates are replaced by controlled-$\sqrt{X}$ (C-$\sqrt{X}$) gates (see \cref{sec: bias} and \cref{sec:derive_gen_framework}). While the second layer is in fact a layer of controlled-$\sqrt{X}^\dagger$ gates, we refer both of them as C-$\sqrt{X}$ layers for simplicity. This version of H-VEC leaves only $X$ noise behind (rather than $Y$ noise when applying C-H) and effectively filters out all other noise. Hence, looking back at the long-range lattice surgery protocol mentioned in \cref{sec: surgery-background}, by placing the splitting, transmitting and merging steps of the boundary in between the two C-$\sqrt{X}$ layers of H-VEC as shown in \cref{fig: surgery}~(b), \emph{we can remove all of the dangerous $Z$ errors on the boundary qubits that would otherwise have corrupted long-range lattice surgery}. We apply C-$\sqrt{X}$ instead of C-H on the transmitted boundary, since the former commutes with both the $Z$ checks on the bulk during splitting and the full set of $X$ checks during merging, so that H-VEC does not interfere with the original long-range lattice surgery process. 

During the splitting and transmitting process, even though the transmitted boundary is in a repetition code state, we do not explicitly perform the parity checks associated with the repetition code. These parities are inferred from the initial parity checks before the split and the parity checks on the bulk during splitting. When merging, only $X$ checks are performed and all $Z$ checks are inferred from existing parity results. All of these further ensure that the splitting, transmitting and merging processes do not interfere with the H-VEC circuit, and also introduces fewer gate noise onto the repetition code.

Of course, these additional C-$\sqrt{X}$ gates also introduce additional noise on the transmitted boundary. Any $X$ errors introduced by these gates would commute right through the H-VEC circuit and get picked up by the full set of stabiliser checks at the very end, and $Z$ errors after the second C-$\sqrt{X}$ are suppressed by the same rounds of checks. Any $Z$ errors occurring between the two C-$\sqrt{X}$ layers are covered by H-VEC, and thus will not corrupt the transmitted boundary. To detect and correct $Z$ errors on the transmitted boundary before the first C-$\sqrt{X}$ layer, we introduce additional rounds of $X$ checks after the first C-$\sqrt{X}$ layer as shown in \cref{fig: surgery}~(b). 

The only remaining errors that we have not accounted for are those on the control qubit. As discussed in \cref{subsec: limits} and \cref{app: robustness}, protocols like H-VEC are inherently robust against many common noise types on the control qubit, which include depolarisation, dephasing and amplitude damping~\cite{linErrormitigatedQuantumMetrology2025}. If there is still some small residual noise, one can apply additional error mitigation techniques, such as PEC~\cite{liu2025virtual} or tomography purification~\cite{huo2022dual} on the control qubit. 

The protocol described here performs a fault-tolerant logical $X\otimes X$ measurement. Similar arguments can be extended to fault-tolerant $Z\otimes Z$ measurements with C-$\sqrt{X}$ replaced by C-$\sqrt{Z}$.

\section{General virtual error correction framework}\label{sec: general}

The scheme presented in \cref{sec: vqc} relies on a non-trivial interplay between QEM and QEC, where by QEC we also include instances that encode qubits using classical codes. In the following, we will present a generalised framework that formalises this type of integration, which utilises a controlled conjugation of the noise channel and a stabiliser measurement.

For a given stabiliser code defined by the code space projector $\Pi_E$, we can always find a set of correctable errors, denoted as $\{E_i\}$, such that the code is non-degenerate. Let us choose the error basis to be the canonical one such that $E_i \Pi_E E_i^\dagger$ is the syndrome subspace corresponding to the syndrome $i$. The code subspace then satisfies the Knill-Laflamme condition for non-degenerate codes~\cite{knill1997theory}:
\begin{align}\label{eqn:kl_cond}
    \Pi_E E_i^\dagger E_j \Pi_E = \lambda_i \delta_{ij} \Pi_E \quad \forall i,j \quad (\lambda_i > 0)
\end{align} 
There is another set of errors $\{F_j\}$ that are undetectable (not just uncorrectable) by $\Pi_E$, which implies that $\{F_j\}$ do not take any code states out of the code space (otherwise it will be detectable)
\begin{align}\label{eqn:undetectable_cond_1}
(I - \Pi_E )F_j \Pi_E = 0\ \ \Rightarrow\ \ \Pi_E F_j \Pi_E = F_j \Pi_E \quad \forall j.
\end{align}
Hence, a noise channel $\mathcal{N}$ of the form
\begin{align}\label{eq: noise}
    \mathcal{N}[\;\cdot\;] = \sum_{i,j}p_{i,j} E_i F_j \,\cdot\, F_j^\dagger E_i^\dagger
\end{align}
cannot be fully corrected by a code state in $\Pi_E$ since $\{F_j\}$ are undetectable. Here $p_{i,j}$ will be a valid distribution when we define $\{E_iF_j\}$ in way such that they are properly normalised.

\begin{figure}
\includegraphics[width=0.38\textwidth]{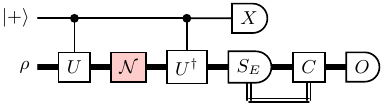}
\caption{Quantum circuit involved in the general VEC framework. The input state $\rho$ is some code state within the code space of $\Pi_E$. By choosing $U$ appropriately given the noise channel $\mathcal{N}$ that we wish to correct, the stabiliser measurement $S_E$, again of $\Pi_E$, effectively projects the noise in the logical register into one that is correctable using the obtained syndrome upon post-processing via the $X$-basis measurement of the control qubit.}\label{fig: general circ}
\end{figure}

Now suppose we manage to find some unitary $U$ that can transform between $\{E_i\}$ and $\{F_j\}$ via
\begin{equation}\label{eq: assumption 1}
    U^\dagger E_iF_jU = \beta_{ij}E_jF_i \quad \forall i,j,
\end{equation}
for some complex number $\beta_{ij}$. This can happen under several practical scenarios in the examples that we have seen and also in some more general cases outlined in  \cref{sec:derive_gen_framework}. If this is the case, by including an additional control qubit and a pair of controlled unitary operations defined by $U$, our framework enables correcting both $\{E_i\}$ and $\{F_j\}$ on an input state encoded in $\Pi_E$ using the circuit shown in \cref{fig: general circ}.

As shown in \cref{sec:derive_gen_framework}, the effective output state of \cref{fig: general circ}, after post-processing on the $X$ measurement outcome and obtaining the syndrome $k$ on the $S_E$ check, is given by
\begin{equation}
    p_{k,k} \lambda_k^2 \bigg[ \mathrm{Re}[\beta_{kk}]E_kF_k\rho F_k^\dagger E_k^\dagger\bigg].
\end{equation}
Therefore, upon obtaining the syndrome $k$, we can simply apply $(E_kF_k)^{-1}$ as the correction operation (denoted $C$ in \cref{fig: general circ}) to undo the errors and recover the code state $\rho$, and the scalar factor can be removed by multiplying $\lambda_k^{-2}\mathrm{Re}[\beta_{kk}]^{-1}$ in classical software to the output of the circuit. What \cref{fig: general circ} and the relevant post-processing is doing is to effectively post-select for the noise elements $\{E_iF_j\}$ that are invariant (up to a scalar factor) under the transformation in \cref{eq: assumption 1}.

Correcting for the projected errors and averaging over all possible projection outcomes, the virtual state obtained right before measuring an observable $O$ is given by $P \rho$, where $P = \sum_k p_{k,k} \leq 1$ is the normalisation factor, which is simply the expectation value of the $X$ measurement on the control qubit. The sampling overhead would thus be given by $P^{-2}$.

\section{Discussion}\label{sec:discussion}

We introduced Hadamard-based virtual error correction (H-VEC), which enables correcting quantum errors using a classical code and one additional qubit. By post-processing the measurement results of the control qubit, the scheme can provide even stronger error suppression power than its constituent classical code and its quantum counterpart. This is demonstrated in the detailed comparison between the repetition code, its H-VEC variant, and the unrotated surface code in \cref{sec:performance}. Since classical codes are simpler than quantum codes in every stage of the error correction pipeline, the scheme may alleviate many of the challenges involved in QEC. We presented a practical use case, in which H-VEC is used to quadratically reduce the required number of qubits to teleport in long-range lattice surgery, while ensuring fault-tolerance throughout the process. We further generalised our scheme into a broader VEC framework, which describes a non-trivial interplay between QEM and QEC. In the Appendices, we also show that our scheme can be particularly effective under biased noise, can be flexibly integrated into various hardware architectures, and has interesting implications in entanglement purification.

Our protocol is applicable to any classical code, and it would be interesting to investigate the exact performance of our protocol beyond the repetition code, considering variations in connectivity requirements, encoding rate, the availability of efficient decoding algorithms, etc. Interesting examples may be high-rate codes such as the Hamming code, or any other classical codes without an efficient quantum counterpart. Such investigations would be particularly interesting for extending our efficient long-range lattice surgery protocols to more general quantum codes beyond surface codes.

To study how specific classical codes may be implemented on real hardware is also important, and we described one such consideration in \cref{app: transversal}. As the connectivity required by classical codes are generally significantly simpler than their quantum counterparts, one may study the compatibility between specific combinations of classical codes and hardware architectures, including modular devices that apply inter-module syndrome extraction. Considerations regarding how to allocate the control qubit would be crucial to this end, including the search for alternative circuits that render the same virtual code construction.

Finally, one could use the general VEC framework discussed in \cref{sec: general} to discover additional schemes that augment the power of existing error correction protocols, possibly with an input state encoded in a quantum code instead of a classical code. Our work provides a novel and resource-efficient pathway for suppressing errors, redefining the trade-offs between physical hardware, circuit depth, and classical processing. It will thus be important to look into such trade-offs in different application scenarios, such as the 2-to-1 virtual EPP described in \cref{app: magic} and the error-suppression of magic states.

\section*{Acknowledgements}

Numerical simulations utilise the Quantum Exact Simulation Toolkit (QuEST)~\cite{jones2019quest} via the QuESTlink~\cite{jones2020questlink} frontend, as well as Stim~\cite{gidney2021stim}. Quantum circuits are illustrated using the Quantikz LaTeX package~\cite{kay2023tutorial}. TA acknowledges support by the Oxford-Uehiro Graduate Scholarship Programme. JFG thanks the UK Engineering and Physical Sciences Research Council (EPSRC) for funding under the Horizon Europe Guarantee (EP/Y026438/1), and the European Research Council (ERC) for selecting and approving the proposal (ERC Starting Grant: MICRON-QC - 101077098). ZC acknowledges support from the EPSRC project Robust and Reliable Quantum Computing (RoaRQ, EP/W032635/1) and the EPSRC quantum technologies career acceleration fellowship (UKRI1226). JFG is a director of Quantum Fabrix Ltd. (QFX). QFX was not involved in directly funding this
work. For the purpose of Open Access, the author has applied a CC BY public copyright licence
to any Author Accepted Manuscript version arising from this submission.

\appendix
\crefalias{section}{appendix}

\section{H-VEC Derivation}\label{sec:derivation}
In order to correct both $X$ and $Z$ noise with the input bit-flip code state $\rho_Z$, we employ the circuit in \cref{fig: vqc}. Before the $X$ measurement of the control qubit, this circuit outputs the state
\begin{align}\label{eq: rho_full1}
\begin{split}
    \rho_{\text{full},1} &= \frac{1}{2}\sum_{\vec{x}, \vec{{z}} \in \mathbb{E}_{\mathrm{cor}}}p_{\vec{x}, \vec{z}}  \bigg[ \ketbra{0} \otimes X^{\vec{x}} Z^{\vec{z}} \rho_Z Z^{\vec{z}} X^{\vec{x}} \\
   &\quad + \ketbra{1} \otimes Z^{\vec{x}} X^{\vec{z}} \rho_Z  X^{\vec{z}}Z^{\vec{x}} \\
   &\quad + \ketbra{0}{1} \otimes X^{\vec{x}} Z^{\vec{z}} \rho_Z  X^{\vec{z}} Z^{\vec{x}}  + c.c. \bigg]\\
   & = \frac{1}{2}\sum_{\vec{x}, \vec{{z}} \in \mathbb{E}_{\mathrm{cor}}}p_{\vec{x}, \vec{z}}  \bigg[ \ketbra{0} \otimes X^{\vec{x}} Z^{\vec{z}} \rho_Z Z^{\vec{z}} X^{\vec{x}}  \\
   &\quad + \ketbra{1} \otimes X^{\vec{z}} Z^{\vec{x}}  \rho_Z  Z^{\vec{x}} X^{\vec{z}} \\
   &\quad + (-1)^{\vec{x} \cdot \vec{z}} \ketbra{0}{1} \otimes X^{\vec{x}} Z^{\vec{z}} \rho_Z   Z^{\vec{x}}X^{\vec{z}}  + c.c. \bigg],
\end{split}
\end{align}
where the last c.c. term is the complex conjugate of the third term. 

By performing stabiliser measurements on the main register and obtaining the syndrome $\vec{k}$, the output state is
\begin{align}\label{eq: rho_k1}
\begin{split}
    \rho_{\vec{k},1} &= \frac{1}{P_{\vec{k}}}X^{\vec{k}}\Pi_{Z}X^{\vec{k}}\rho_{\text{full},1}X^{\vec{k}}\Pi_{Z}X^{\vec{k}} \\
    &= \frac{1}{2P_{\vec{k}}}\sum_{\vec{x}, \vec{{z}} \in \mathbb{E}_{\mathrm{cor}}}p_{\vec{x}, \vec{z}}  \bigg[ \ketbra{0} \otimes \delta_{\vec{x}, \vec{k}} X^{\vec{k}}  Z^{\vec{z}} \rho_Z Z^{\vec{z}} X^{\vec{k}}  \\
    &\quad + \ketbra{1} \otimes \delta_{\vec{z}, \vec{k}} X^{\vec{k}} Z^{\vec{x}}  \rho_Z  Z^{\vec{x}} X^{\vec{k}} \\
   &\quad+ (-1)^{\vec{x} \cdot \vec{z}} \ketbra{0}{1} \otimes \delta_{\vec{x}, \vec{k}} \delta_{\vec{z}, \vec{k}}X^{\vec{k}} Z^{\vec{z}} \rho_Z   Z^{\vec{x}}X^{\vec{k}}  + c.c. \bigg]\\
   & = \frac{1}{2P_{\vec{k}}} \bigg[ \sum_{\vec{{z}} \in \mathbb{E}_{\mathrm{cor}}} p_{\vec{k}, \vec{z}} \ketbra{0} \otimes X^{\vec{k}}  Z^{\vec{z}} \rho_Z Z^{\vec{z}} X^{\vec{k}}\\
   &\quad + \sum_{\vec{x} \in \mathbb{E}_{\mathrm{cor}}} p_{\vec{x},\vec{k}} \ketbra{1} \otimes X^{\vec{k}} Z^{\vec{x}}  \rho_Z  Z^{\vec{x}} X^{\vec{k}}\\
   &\quad + p_{\vec{k}, \vec{k}}(-1)^{\abs{\vec{k}}} \ketbra{0}{1} \otimes Y^{\vec{k}} \rho_Z   Y^{\vec{k}}  + c.c. \bigg],
\end{split}
\end{align}
where we applied the Knill-Laflamme condition stated in \cref{eqn:kl_classical} and the fact that $Z$ errors commute with $\Pi_Z$, similarly as in \cref{eq: general projection} for the general case. The normalisation factor $P_{\vec{k}}$ corresponds to the probability of measuring the $\vec{k}$th syndrome, and it ensures that $\mathrm{Tr}(\rho_{\vec{k},1}) = 1$. It is given by
\begin{align*}
    P_{\vec{k}} &= \frac{1}{2}\mathrm{Tr}\bigg( \sum_{\vec{{z}} \in \mathbb{E}_{\mathrm{cor}}} p_{\vec{k}, \vec{z}} \ketbra{0} \otimes X^{\vec{k}}  Z^{\vec{z}} \rho_Z Z^{\vec{z}} X^{\vec{k}} \\
    &\quad + \sum_{\vec{x} \in \mathbb{E}_{\mathrm{cor}}} p_{\vec{x},\vec{k}} \ketbra{1} \otimes X^{\vec{k}} Z^{\vec{x}}  \rho_Z  Z^{\vec{x}} X^{\vec{k}} \\
   &\quad + p_{\vec{k}, \vec{k}} (-1)^{\abs{\vec{k}}} \ketbra{0}{1} \otimes X^{\vec{k}} Z^{\vec{k}} \rho_Z   Z^{\vec{k}}X^{\vec{k}}  + c.c. \bigg)\\
   & = \frac{1}{2} \sum_{\vec{{z}} \in \mathbb{E}_{\mathrm{cor}}} (p_{\vec{k}, \vec{z}} + p_{\vec{z}, \vec{k}}).
\end{align*}
Notice that, since we know the syndrome $\vec{k}$, we can restore $\rho_Z$ in the last two terms of $\rho_{\vec{k},1}$ by applying the $Y^{\vec{k}}$ correction and the phase $(-1)^{\abs{\vec{k}}}$ to the output state $\rho_{\vec{k},1}$. This leads to
\begin{align}\label{eq: rho_k2}
\begin{split}
    \rho_{\vec{k},2} & = (-1)^{\abs{\vec{k}}} Y^{\vec{k}}\rho_{\vec{k},1}Y^{\vec{k}}\\
    & = \frac{1}{2P_{\vec{k}}} \bigg[ (-1)^{\abs{\vec{k}}}\ketbra{0} \otimes \left(\sum_{\vec{{z}} \in \mathbb{E}_{\mathrm{cor}}} p_{\vec{k}, \vec{z}} Z^{\vec{z} + \vec{k}} \rho_Z Z^{\vec{z} + \vec{k}}\right) \\
    &\quad +  (-1)^{\abs{\vec{k}}}\ketbra{1} \otimes \left(\sum_{\vec{x} \in \mathbb{E}_{\mathrm{cor}}} p_{\vec{x},\vec{k}} Z^{\vec{x} + \vec{k}}  \rho_Z  Z^{\vec{x} + \vec{k}}\right) \\
    &\quad+ p_{\vec{k}, \vec{k}}\ketbra{0}{1} \otimes \rho_Z  + c.c. \bigg].
\end{split}
\end{align}
Thus, the mixture of all possible $\vec{k}$ outcomes is given by
\begin{align*}
    \rho_{\mathrm{full},2} & = \sum_{\vec{k}\in \mathbb{E}_{\mathrm{cor}}} P_{\vec{k}} \rho_{\vec{k},2}\\
    & = \frac{1}{2} \bigg[ \ketbra{0} \otimes \left(\sum_{\vec{{z}},\vec{k} \in \mathbb{E}_{\mathrm{cor}}} (-1)^{\abs{\vec{k}}} p_{\vec{k}, \vec{z}} Z^{\vec{z} + \vec{k}} \rho_Z Z^{\vec{z} + \vec{k}}\right) \\
    &\quad +  \ketbra{1} \otimes \left(\sum_{\vec{x},\vec{k} \in \mathbb{E}_{\mathrm{cor}}} (-1)^{\abs{\vec{k}}} p_{\vec{x},\vec{k}} Z^{\vec{x} + \vec{k}}  \rho_Z  Z^{\vec{x} + \vec{k}}\right) \\
    &\quad + \sum_{\vec{k}\in \mathbb{E}_{\mathrm{cor}}} p_{\vec{k}, \vec{k}} \ketbra{0}{1} \otimes \rho_Z  + c.c. \bigg].
\end{align*}
We see that only $Z$ errors are left on the first two terms, which is as expected since our code should be able to correct all incoming $X$ errors. For the last two terms however, all of the errors are removed! In order to extract information from the last two terms, we can perform an $X$ measurement on the control qubit to remove the first two terms. More exactly, to obtain $\Tr(O \rho_Z)$, we can apply the ratio
\begin{align*}
     \Tr(O \rho_Z) = \frac{\Tr((X \otimes O) \rho_{\mathrm{full},2})}{\Tr((X \otimes I) \rho_{\mathrm{full},2})} = \frac{\expval{X \otimes O}}{\expval{X \otimes I}}
\end{align*}
Here, $\expval{X \otimes O}$ and $\expval{X \otimes I}$ follow the same notation as in the main text, denoting the expectation values measured from the circuit in \cref{fig: vqc}. The normalisation factor is given by
\begin{align*}
  \Tr((X \otimes I) \rho_{\mathrm{full},2}) = \sum_{\vec{k}\in \mathbb{E}_{\mathrm{cor}}} p_{\vec{k}, \vec{k}} 
\end{align*}

\section{Uncorrectable Errors in H-VEC} \label{sec:uncorrect_error}
For a given classical bit-flip code, let us denote the set of supports for its $X$ logical operators as $\mathbb{X}$, i.e., $X^{\vec{\ell}}$ is a logical operator for all $\vec{\ell} \in \mathbb{X}$. Note that for classical codes there is no equivalence between different logical operators, so for a code with $k$ logical qubits, we will have $\abs{\mathbb{X}} = 2^k$. The analogue of the Knill-Laflamme condition for all possible errors, i.e., including uncorrectable ones, is given by
\begin{align}\label{generalised_KL}
    \Pi_Z X^{\vec{u}}X^{\vec{v}} \Pi_Z = \begin{cases}
        X^{\vec{u} \oplus \vec{v}} \Pi_Z \quad &\text{if } \vec{u} \oplus \vec{v} \in \mathbb{X}\\
        0 \quad &\text{otherwise}
    \end{cases}
\end{align}
where $\oplus$ denotes a bitwise XOR. If $\vec{u}$ and $\vec{v}$ are both in the correctable set $\mathbb{E}_{\mathrm{cor}}$, then $\vec{u} \oplus \vec{v} \in \mathbb{X}$ implies that $\vec{u} = \vec{v}$, and we recover \cref{eqn:kl_classical}.

An important feature of H-VEC is that the stabiliser measurement is applied to $X^{\vec{x}}$ on one side of $\rho_Z$ and $X^{\vec{z}}$ on the other, which can be seen in \cref{eq: rho_k1}, \cref{eq: general projection}, and below. As we will see, \cref{generalised_KL} implies that a stabiliser measurement that projects into the $X^{\vec{k}}\Pi_{Z}X^{\vec{k}}$ syndrome subspace only keeps terms where both $\vec{x} \oplus \vec{k}$ and $\vec{z} \oplus \vec{k}$ are supports of logical operators and all other terms with different $\vec{x}$ and $\vec{z}$ are removed.

To study residual errors, we apply the same analysis as in \cref{eq: rho_k1}, but now consider all errors including uncorrectable ones. Suppressing terms that involve $\ketbra{0}$ and $\ketbra{1}$, which will not contribute after we perform the $X$-basis measurement of the control qubit, we obtain
\begin{align*}
    \rho_{\vec{k},1}
   & = \frac{1}{2P_{\vec{k}}} \bigg[ ... + \ketbra{0}{1} \otimes \sum_{\vec{x}, \vec{z}} p_{\vec{x}, \vec{z}}(-1)^{\vec{x} \cdot \vec{z}} \\
   &\quad X^{\vec{k}}\Pi_{Z}X^{\vec{k}}X^{\vec{x}} Z^{\vec{z}} \rho_Z   Z^{\vec{x}}X^{\vec{z}} X^{\vec{k}}\Pi_{Z}X^{\vec{k}}  + c.c. \bigg]\\
   & = \frac{1}{2P_{\vec{k}}} \bigg[ ... + \ketbra{0}{1} \otimes \sum_{\vec{x}\oplus \vec{k}, \vec{z}\oplus \vec{k} \in \mathbb{X}} p_{\vec{x}, \vec{z}}(-1)^{\vec{x} \cdot \vec{z}}  \\
   &\quad X^{\vec{x}} Z^{\vec{z}} \rho_Z Z^{\vec{x}}X^{\vec{z}}  + c.c. \bigg]\\
    & = \frac{1}{2P_{\vec{k}}} \bigg[ ... + \ketbra{0}{1} \otimes \sum_{\vec{u}, \vec{v} \in \mathbb{X}} p_{\vec{u} \oplus \vec{k}, \vec{v} \oplus \vec{k}}(-1)^{(\vec{u} \oplus \vec{k}) \cdot (\vec{v} \oplus \vec{k})}  \\
   &\quad X^{\vec{u} \oplus \vec{k}} Z^{\vec{v} \oplus \vec{k}} \rho_Z   Z^{\vec{u} \oplus \vec{k}}X^{\vec{v} \oplus \vec{k}}  + c.c. \bigg],
\end{align*}
where we applied \cref{generalised_KL} in the second equality, which keeps only the terms with $\vec{u}, \vec{v} \in \mathbb{X}$ for $\vec{u} = \vec{x} \oplus \vec{k}$ and $\vec{v} = \vec{z} \oplus \vec{k}$. Using $Z^{\vec{u} \oplus \vec{k}}X^{\vec{v} \oplus \vec{k}} = (-1)^{\vec{k} \cdot\vec{v}}Z^{\vec{u}}X^{\vec{v}}Z^{\vec{k}}X^{\vec{k}}$, and similarly for $X^{\vec{u} \oplus \vec{k}}Z^{\vec{v} \oplus \vec{k}}$, and also expanding the product in the power of $(-1)^{(\vec{u} \oplus \vec{k}) \cdot (\vec{v} \oplus \vec{k})}$, we have
\begin{align*}
    \rho_{\vec{k},1}
    & = \frac{1}{2P_{\vec{k}}} \bigg[ ... + \ketbra{0}{1} \otimes \sum_{\vec{u}, \vec{v} \in \mathbb{X}} p_{\vec{u} \oplus \vec{k}, \vec{v} \oplus \vec{k}}(-1)^{\vec{u} \cdot \vec{v}} (-1)^{\abs{\vec{k}}}  \\
   &\quad Y^{\vec{k}}X^{\vec{u}} Z^{\vec{v}} \rho_Z   Z^{\vec{u}}X^{\vec{v}}Y^{\vec{k}}  + c.c. \bigg]
\end{align*}
Applying our $Y^{\vec{k}}$ correction and the phase $(-1)^{\abs{\vec{k}}}$, we then have
\begin{align*}
    \rho_{\vec{k},2}
    & = \frac{1}{2P_{\vec{k}}} \bigg[ ... + \ketbra{0}{1} \otimes\\ 
    &\sum_{\vec{u}, \vec{v} \in \mathbb{X}} p_{\vec{u} \oplus \vec{k}, \vec{v} \oplus \vec{k}}  X^{\vec{u}} Z^{\vec{v}} \rho_Z   X^{\vec{v}} Z^{\vec{u}}  + c.c. \bigg].
\end{align*}
The mixture of all possible $\vec{k}$ outcomes is given by
\begin{align*}
    &\rho_{\mathrm{full},2}  = \sum_{\vec{k}\in \mathbb{E}_{\mathrm{cor}}} P_{\vec{k}} \rho_{\vec{k},2}\\
    & = \frac{1}{2} \bigg[ ... +  \ketbra{0}{1} \otimes  \\
    & \sum_{\vec{k}\in \mathbb{E}_{\mathrm{cor}}} \sum_{\vec{u}, \vec{v} \in \mathbb{X}} p_{\vec{u} \oplus \vec{k}, \vec{v} \oplus \vec{k}} X^{\vec{u}} Z^{\vec{v}} \rho_Z   X^{\vec{v}} Z^{\vec{u}}  + c.c. \bigg]
\end{align*}
The normalisation constant is obtained by the expectation value of the $X$ measurement on the control qubit, which removes all of the terms contained in the suppressed terms (in ...) above, and gives
\begin{align*}
    &P_{\mathrm{full}} = \Tr(\left(X\otimes I\right)\rho_{\mathrm{full},2}) \\
    &= \sum_{\vec{k}\in \mathbb{E}_{\mathrm{cor}}} \sum_{\vec{u}, \vec{v} \in \mathbb{X}} p_{\vec{u} \oplus \vec{k}, \vec{v} \oplus \vec{k}} \Re{\Tr(X^{\vec{u}} Z^{\vec{v}} \rho_Z   X^{\vec{v}} Z^{\vec{u}})}\\
    & = \sum_{\vec{k}\in \mathbb{E}_{\mathrm{cor}}} \sum_{\vec{u}, \vec{v} \in \mathbb{X}} p_{\vec{u} \oplus \vec{k}, \vec{v} \oplus \vec{k}} \Re{(-1)^{\abs{\vec{v}}} i^{\abs{\vec{u} \oplus \vec{v}}}\Tr(Y^{\vec{u} \oplus \vec{v}}\rho_Z)}
\end{align*}
where we used
\begin{align*}
X^{\vec{v}} Z^{\vec{u}}X^{\vec{u}} Z^{\vec{v}}
     & = (-1)^{\abs{\vec{v}}} i^{\abs{\vec{u} \oplus \vec{v}}} Y^{\vec{u} \oplus \vec{v}}. 
\end{align*}
The correctable terms are those where $\vec{x} = \vec{z} = \vec{k}$ or, equivalently, $\vec{v} = \vec{u} = \vec{0}$. This component within $P_{\mathrm{full}}$ is given by
\begin{align*}
    P_{\mathrm{cor}} 
    & = \sum_{\vec{k} \in \mathbb{E}_{\mathrm{cor}}}  p_{\vec{k}, \vec{k}},
\end{align*}
which is the same as the normalisation factor obtained in \cref{sec:derivation}. This is because codes are usually chosen such that uncorrectable errors occur with small probabilities, and thus $P_{\mathrm{full}}$ can be approximated using $ P_{\mathrm{cor}}$ when trying to calculate the sampling overhead.

At this point, it is also worth noting that when assuming that uncorrectable errors occur with vanishingly small probabilities, as was effectively the case in \cref{sec:derivation}, the virtual state is exactly protected even if we do not apply the $(-1)^{|\vec{k}|}$ phase correction. However, its presence leads to an increase of $P_{\mathrm{full}}$, which both amplifies residual logical errors and increases the sampling cost, so its correction is crucial in practice.

The effective logical error rate is approximately (the exact bias will depend on the observable of interest)
\begin{align*}
   P_{\mathrm{L}} \approx \frac{P_{\mathrm{full}} - P_{\mathrm{cor}}}{P_{\mathrm{full}}}
\end{align*}
Since $\Re{(-1)^{\abs{\vec{v}}} i^{\abs{\vec{u} \oplus \vec{v}}}\Tr(Y^{\vec{u} \oplus \vec{v}}\rho_Z)} \leq 1$, we can define an upper bound of $P_{\mathrm{full}}$ as
\begin{align}
    P_{\mathrm{full}} \leq P_{\mathrm{max}} &= \sum_{\vec{k}\in \mathbb{E}_{\mathrm{cor}}} \sum_{\vec{u}, \vec{v} \in \mathbb{X}} p_{\vec{u} \oplus \vec{k}, \vec{v} \oplus \vec{k}}\\
    & = \sum_{\vec{v} \in \mathbb{X}} \sum_{\vec{x}}  p_{\vec{x}, \vec{x} \oplus \vec{v}}
\end{align}
where we used $\sum_{\vec{k}\in \mathbb{E}_{\mathrm{cor}}} \sum_{\vec{u} \in \mathbb{X}} f(\vec{u} \oplus \vec{k}) = \sum_{\vec{x}}f(\vec{x})$ with no restriction on $\vec{x}$. Note the absence of the probabilities of many uncorrectable components here, since we have essentially kept only those $p_{\vec{x}, \vec{z}}$ with $\vec{x}$ and $\vec{z}$ differing by a logical operator, and all other error components are removed, even if uncorrectable. We can use $P_{\mathrm{max}}$ to build a rough upper bound on the effective logical error rate as
\begin{align*}
   P_{\mathrm{L}} \lesssim \frac{P_{\mathrm{max}} - P_{\mathrm{cor}}}{P_{\mathrm{max}}}\lesssim \frac{P_{\mathrm{max}} - P_{\mathrm{cor}}}{P_{\mathrm{cor}}}
\end{align*}

\section{Derivation of the Performances of Representative Example Codes}\label{sec:example_derivation}
\subsection{Bit-Flip Repetition Code} 
The probability of logical phase-flip errors is given by
\begin{align*}
    p_{\textrm{L,rep}, Z} &= \sum_{w = 1}^{d} \binom{d}{w} \left(1-\frac{2p}{3}\right)^{d-w} \left(\frac{2p}{3}\right)^{w}\\
    & = 1 - \left(1-\frac{2p}{3}\right)^d ,
\end{align*}
which is simply $1$ minus the probability of pure $X$ errors. Thus, at small $p$, we have
\begin{align*}
    p_{\textrm{L,rep}, Z} 
    & \approx \frac{2dp}{3}.
\end{align*}

The probability of logical bit-flip errors is given by:
\begin{align*}
    p_{\textrm{L,rep}, X} &= \sum_{w =(d+1)/2}^{d} \binom{d}{w} \left(1-\frac{2p}{3}\right)^{d-w} \left(\frac{2p}{3}\right)^{w}.
\end{align*}
At small $p$, only the leading term remains, and we have
\begin{align}
    p_{\textrm{L,rep}, X} &\approx \binom{d}{(d+1)/2} \left(1-\frac{2p}{3}\right)^{\frac{d-1}{2}} \left(\frac{2p}{3}\right)^{\frac{d+1}{2}}.
\end{align}

\subsection{Virtual Quantum Repetition Code} 

There is just one $X$ logical operator for the repetition code, which is $X^{\vec{1}}$. Thus, the set of logical operators for the repetition code is $\mathbb{X} = \{\vec{0}, \vec{1}\}$. 

Following arguments in \cref{sec:uncorrect_error}, the full magnitude of the remaining terms after H-VEC is
\begin{align*}
    P_{\mathrm{full}} & = \sum_{\vec{v} \in \{\vec{0}, \vec{1}\}} \sum_{\vec{k}}  p_{\vec{k}, \vec{k} \oplus \vec{v}}\\
    & = \sum_{\vec{k}}  p_{\vec{k}, \vec{k}} + \sum_{\vec{k}}  p_{\vec{k}, \vec{k} \oplus \vec{1}}
\end{align*}
Recall that $p_{\vec{x}, \vec{z}}$ is the probability that the error $X^{\vec{x}}Z^{\vec{z}}$ occurs, such that $p_{\vec{k}, \vec{k}}$ corresponds to the probability that $Y^{\vec{k}}$ occurs. This means
\begin{align*}
    p_{\vec{k}, \vec{k}} = (1-p)^{d-\abs{\vec{k}}} \left(\frac{p}{3}\right)^{\abs{\vec{k}}}. 
\end{align*}
Hence, with $\abs{\vec{k}} = w$, we have
\begin{align*}
    \sum_{\vec{k}}  p_{\vec{k}, \vec{k}} &= \sum_{w = 0}^{d} \binom{d}{w}  (1-p)^{d-w} \left(\frac{p}{3}\right)^{w}\\
    & = (1-p + \frac{p}{3})^{d} = \left(1-\frac{2p}{3}\right)^{d}.
\end{align*}
This makes sense since $(1-2p/3)^{d}$ is indeed the probability that no position takes $X$ or $Z$, which is the probability that only pure $Y$ errors occur.

On the other hand, $p_{\vec{k}, \vec{k} \oplus \vec{1}}$ means that we have $\abs{\vec{k}}$ positions with $X$ errors occurring and $Z$ errors occurring in the rest of the positions. Hence, with each position, we will choose $X$ or $Z$, and the total probability of such strings is
\begin{align*}
    \sum_{\vec{k}}  p_{\vec{k}, \vec{k} \oplus \vec{1}} = \left(\frac{2p}{3}\right)^d.
\end{align*}
Therefore, we have
\begin{align*}
    P_{\mathrm{max}} & =\left(1-\frac{2p}{3}\right)^{d} + \left(\frac{2p}{3}\right)^d.
\end{align*}

The magnitude of correctable errors is simply the sum of the probabilities of all $Y$ errors up to weight $(d-1)/2$
\begin{align}\label{eqn:p_corr}
    P_{\mathrm{cor}} & =\sum_{w = 0}^{\frac{d-1}{2}} \binom{d}{w}  (1-p)^{d-w} \left(\frac{p}{3}\right)^{w}.
\end{align}
The logical error rate is thus given by
\begin{align}
    p_{\textrm{L,vec}} \lesssim \frac{P_{\mathrm{max}} - P_{\mathrm{cor}}}{ P_{\mathrm{max}}}.
\end{align}
For small $p$, we have $ P_{\mathrm{max}} \approx \left(1-2p/3\right)^{d}$, so that
\begin{align*}
    P_{\mathrm{max}} - P_{\mathrm{cor}} &\approx \sum_{w = (d+1)/2}^{d} \binom{d}{w}  (1-p)^{d-w} \left(\frac{p}{3}\right)^{w}\\
    &\approx \binom{d}{(d+1)/2}  (1-p)^{\frac{d-1}{2}} \left(\frac{p}{3}\right)^{\frac{d+1}{2}},
\end{align*}
which implies that
\begin{align}
    p_{\textrm{L,vec}} \approx \left(1-\frac{2p}{3}\right)^{-d} \binom{d}{(d+1)/2}  (1-p)^{\frac{d-1}{2}} \left(\frac{p}{3}\right)^{\frac{d+1}{2}}.
\end{align}

As mentioned in \cref{sec:uncorrect_error}, the sampling overhead can be approximated using the probability of pure \emph{correctable} $Y$ errors occurring, which is given by \cref{eqn:p_corr} and can be approximated by 
\begin{align*}
    P_{Y} \approx \left(1 - \frac{2p}{3}\right)^{d} \quad\text{for small $p$}.
\end{align*}
The approximation is done by taking the whole sum up to $w=d$ for small $p$, which gives the full probability of pure $Y$ errors rather than just the correctable ones. Hence, the sampling overhead factor is approximately
\begin{align*}
    C_{Y} &= P_{Y}^{-2} \approx \left(1 - \frac{2p}{3}\right)^{-2d}.
\end{align*}

\subsection{Surface Code} 
We will focus on the unrotated surface code of \emph{odd distance} $d$ here. The code can correct all errors up to weight $(d-1)/2$, so we will focus on the leading order weight-$(d+1)/2$ errors. However, not all weight-$(d+1)/2$ errors lead to logical errors. They lead to logical errors in the following two cases:
\begin{itemize}
    \item Weight-$(d+1)/2$ errors live on the support of weight-$d$ logical operators. In this case, the complement is weight-$(d-1)/2$, which is smaller than $(d+1)/2$ and thus will be chosen as the correction, leading to logical errors. For the unrotated surface code, there are $d$ such weight-$d$ $X$ logicals running as straight lines between $X$ boundaries. Similarly for $Z$. On each weight-$d$ logical, there are $\binom{d}{(d+1)/2}$ ways that a weight-$(d+1)/2$ error can happen.
    \item Weight-$(d+1)/2$ errors live on the support of weight-$(d+1)$ logical operators. In this case, the complement is also weight-$(d+1)/2$ and thus we will have $0.5$ chance of choosing the wrong correction and lead to logical errors. For unrotated surface code, a weight-$(d+1)$ logical can be obtained by composing a weight-$d$ logical with a weight-$3$ stabiliser of the same type with one-qubit overlap. For the weight-$d$ logicals at the two boundaries, there are $2$ weight-$3$ to choose from for each of them. For the weight-$d$ logicals not at the two boundaries, there are $4$ weight-$3$ to choose from for each of them. So in total, there are $2 \times 2 + (d-2) \times 4 = 4(d-1)$ weight-$(d+1)$ logical $X$ operators and similarly for $Z$. On each weight-$(d+1)$ logical, there are $\binom{d+1}{(d+1)/2}$ ways that a weight-$(d+1)/2$ error can happen.
\end{itemize}
Hence, the total number of ways that a weight-$(d+1)/2$ error can lead to logical errors is
\begin{align*}
    A_{d} &= d \binom{d}{(d+1)/2} + 0.5 \times 4(d-1) \binom{d+1}{(d+1)/2}\\
    & = \left(5d-4\right) \binom{d}{(d+1)/2},
\end{align*}
where we used the identity $\binom{d+1}{(d+1)/2} = 2\binom{d}{(d+1)/2}$. Note that such a count is \emph{not exact} since a given weight-$d$ logical operator can have large overlap with another weight-$d+1$ logical operator and we have not taken this into account. It also depends on how the decoder handles the errors with ambiguous correction possibilities.

The surface code has the same error correction power for both $X$ and $Z$. The probability of the $X$ or $Z$ logical error, keeping only the leading order, is given by:
\begin{align*}
    p_{\textrm{L,sur}} &\approx A_{d} \left(1-\frac{2p}{3}\right)^{\frac{d-1}{2}} \left(\frac{2p}{3}\right)^{\frac{d+1}{2}}\\
    &=  \left(5d-4\right) \binom{d}{(d+1)/2} \left(1-\frac{2p}{3}\right)^{\frac{d-1}{2}} \left(\frac{2p}{3}\right)^{\frac{d+1}{2}}.
\end{align*}

\begin{figure*}
    \centering
    \includegraphics[width=0.9\textwidth]{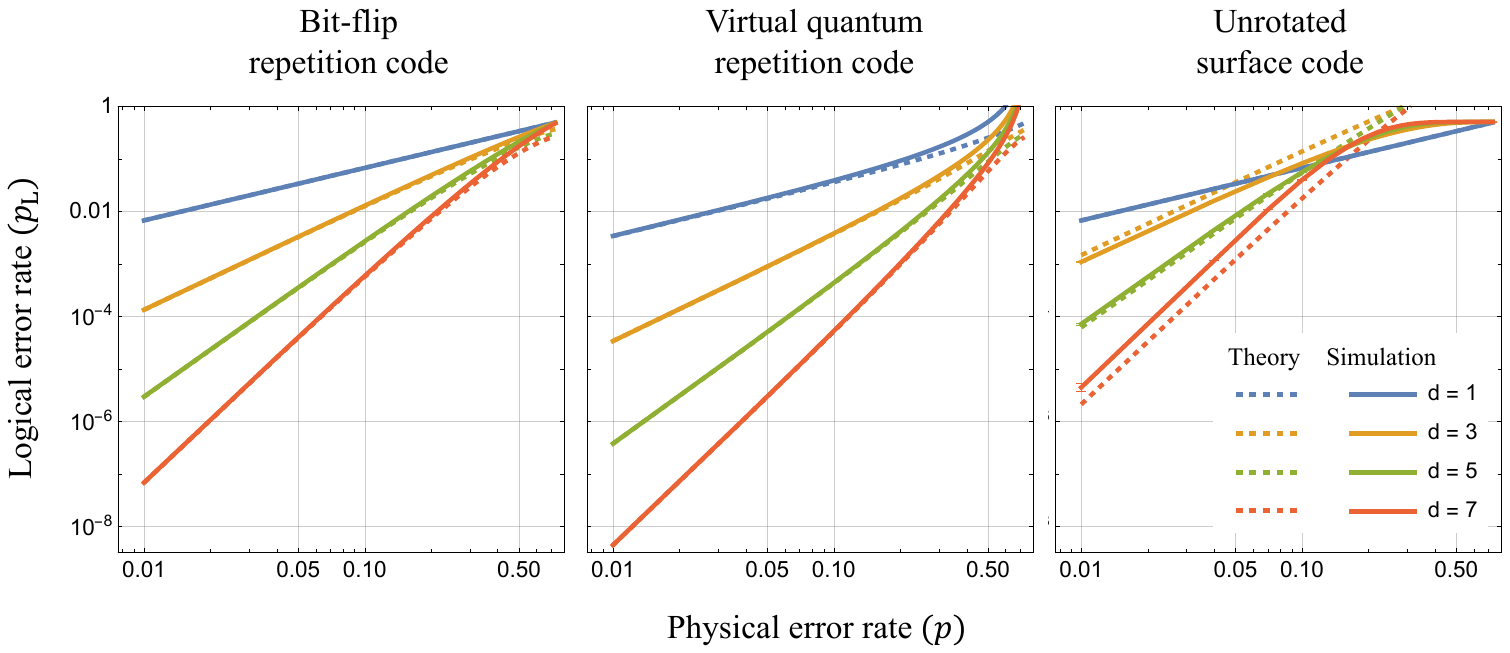}
    \caption{A comparison between simulated and leading-order analytical expression (labelled as ``Theory'') of logical error rate $p_{\mathrm{L}}$ in the logical $Z$-basis, as a function of physical error rate $p$ and code distance $d$. This is shown for the bit-flip repetition code (left), virtual quantum repetition code (middle), and unrotated surface code (right).}
    \label{fig: threshold-theory}
\end{figure*}

\begin{figure}
    \centering
    \includegraphics[width=0.48\textwidth]{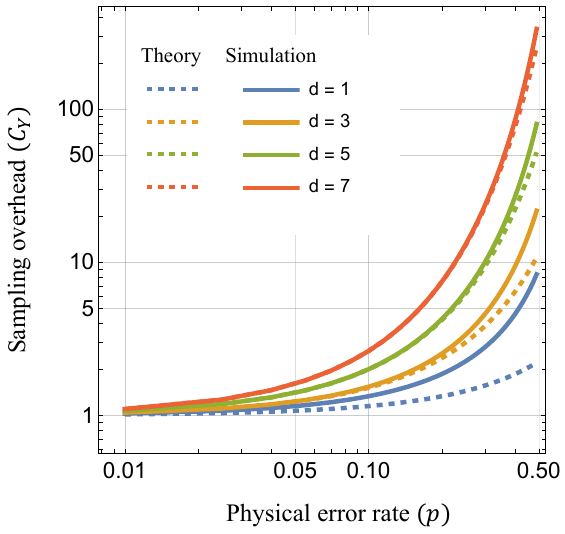}
    \caption{Sampling overhead $C_Y$ of the virtual quantum repetition code as a function of physical error rate $p$ and code distance $d$, plotting both simulated value and leading-order analytical expression (labelled as ``Theory'').}
    \label{fig: sampling-cost}
\end{figure}

\section{Robustness to Common Noise Channels Involving the Control Qubit}\label{app: robustness}
H-VEC suppresses noise in the logical register, but in practice, the additional qubit and gates that enable H-VEC are also noisy. It is thus natural to ask whether these additional noise channels may hinder the efficient removal of bias promised by the technique. Here, we show that the performance of H-VEC is naturally unaffected by many typical noise channels, including amplitude damping and dephasing errors on the control qubit, and depolarising errors of arbitrary weight that involve the control qubit. While not investigated here, these are only examples of error channels that do not introduce noise bias, and the class of errors that fall under this category is more general.

To show this for the amplitude damping channel, let us directly compute the expectation value $\langle X\otimes O\rangle$, noting that the error channel can be expressed using the two Kraus operators $K_0 = \ketbra{0} + \sqrt{1-p} \ketbra{1}$ and $K_1 = \sqrt{p}\ketbra{0}{1}$ for some probability $p$. When the noise is applied before (after) the first (second) C-H layer, the expectation value is given by
\begin{align}
\begin{split}\label{eq: amp_damp}
    &\langle X\otimes O\rangle\\
    =& \sum_{i=0}^1 \mathrm{Tr}\big{(} (X\otimes O) (K_i\otimes I) (\ketbra{+}\otimes \rho_Z)(K_i^\dagger\otimes I)\big{)}\\
    =& \sqrt{1-p} \, \mathrm{Tr}(O \rho_Z).
\end{split}
\end{align}
Thus, we obtain the ideal expectation value as we take the ratio $\langle X\otimes O\rangle / \langle X\otimes I\rangle$. A similar analysis shows that \cref{eq: amp_damp} remains valid when the error occurs between the two C-H layers.

Let us denote a dephasing error channel as inducing the mapping  $\rho \rightarrow \eta\rho + (1-\eta) Z\rho Z$ for some $0 < \eta < 1$. This channel similarly leaves the ratio invariant, since
\begin{align}
\begin{split}
    &\langle X\otimes O\rangle = \eta \mathrm{Tr}\big{(} (X\otimes O) (\ketbra{+}\otimes \rho_Z)\big{)}\\
    &\quad + (1-\eta) \mathrm{Tr}\big{(} (X\otimes O) (Z\otimes I) (\ketbra{+}\otimes \rho_Z)(Z\otimes I)\big{)}\\
    =& [\eta - (1-\eta)] \mathrm{Tr}(O \rho_Z)\\
    =& (2\eta - 1) \mathrm{Tr}(O \rho_Z),
\end{split}
\end{align}
and this does not depend on whether the error occurred between or outside of the two C-H layers since the dephasing channel commutes with the control side of C-H gates.

Finally, consider an error channel that induces the mapping $\rho \rightarrow \eta\rho + (1-\eta) I \otimes \rho_{\mathrm{err}}$, where $I$ is defined on the Hilbert space of the control qubit and $\rho_{\mathrm{err}}$ is an arbitrary density operator corresponding to an erroneous contribution defined on the Hilbert space of the code state. This error channel includes depolarising errors of arbitrary weight that involves the control qubit, and the expectation value under this class of errors is
\begin{align}
\begin{split}
    \langle X\otimes O\rangle =& \eta \mathrm{Tr}\big{(} (X\otimes O) (\ketbra{+}\otimes \rho_Z)\big{)}\\
    &+ (1-\eta) \mathrm{Tr}\big{(} (X\otimes O) (I\otimes \rho_{\mathrm{err}})\big{)}\\
    =& \eta \mathrm{Tr}(O \rho_Z) + (1-\eta) \mathrm{Tr}(X) \mathrm{Tr}(O\rho_{\mathrm{err}})\\
    =& \eta \mathrm{Tr}(O \rho_Z).
\end{split}
\end{align}
The above result holds regardless of when the error channel occurred, since applying a C-H layer to a state in the image of the error mapping does not move it out of this image.

\section{Repeated Checks for Devices with Biased Noise}\label{app: repeat}
To go beyond the code-capacity error model and include errors that arise during stabiliser measurements, it is important to consider whether the assumptions of H-VEC remain valid. Unfortunately, the errors on data qubits that arise during this process would occur after the second C-H gate layer, so our logical register is unprotected from those in general. As we argued in the main text, however, H-VEC is effective when the noise is biased towards $Y$-type errors (see variants of H-VEC for bias towards other Pauli errors in \cref{sec: bias}). Given that stabiliser measurements for classical codes may be performed much faster than their quantum counterparts, we may assume that only $Y$-type errors occur on the data qubits in a biased-noise system during stabiliser measurements, in which case H-VEC remains effective in this scenario.

To see this, let us define the $Y$-type error channel that acts on the logical register between the second C-H layer and the stabiliser measurement by $\mathcal{Y}[\,\cdot\,]=\sum_{\vec{y}\in \mathbb{E}_{\mathrm{cor}}}p_{\vec{y}}Y^{\vec{y}}\cdot Y^{\vec{y}}$ and combined error probabilities $p_{\vec{x},\vec{y},\vec{z}} = p_{\vec{y}}p_{\vec{x},\vec{z}}$. Since there is no hope in correcting errors outside of $\mathbb{E}_{\mathrm{cor}}$, we assume that such errors have vanishingly small probabilities by setting $p_{\vec{x},\vec{y},\vec{z}} = 0$ for all $\vec{x},\vec{y},\vec{z}$ such that $\vec{x}\oplus \vec{y}\notin \mathbb{E}_{\mathrm{cor}}$ or $\vec{z}\oplus \vec{y}\notin \mathbb{E}_{\mathrm{cor}}$. The state before the $X$ measurement corresponds to that in \cref{eq: rho_full1}, but subject to the additional biased error channel as
\begin{align}\label{eq: rho_full_prime}
\begin{split}
    \rho_{\text{full},1}'=& \frac{1}{2} \bigg[...+ \ketbra{0}{1} \otimes \sum_{\vec{x},\vec{y},\vec{z} \in \mathbb{E}_{\mathrm{cor}}} p_{\vec{x},\vec{y},\vec{z}}\\
   &(-1)^{\vec{x} \cdot \vec{z}} Y^{\vec{y}} X^{\vec{x}} Z^{\vec{z}} \rho_Z   Z^{\vec{x}}X^{\vec{z}} Y^{\vec{y}} + c.c. \bigg]\\
   =& \frac{1}{2} \bigg[...+ \ketbra{0}{1} \otimes \sum_{\vec{x},\vec{y},\vec{z} \in \mathbb{E}_{\mathrm{cor}}} p_{\vec{x},\vec{y},\vec{z}}\\
   &(-1)^{\vec{x} \cdot \vec{z}} X^{\vec{x}\oplus \vec{y}} Z^{\vec{z}\oplus \vec{y}} \rho_Z Z^{\vec{x} \oplus \vec{y}}X^{\vec{z}\oplus \vec{y}} + c.c. \bigg],
\end{split}
\end{align}
where we suppressed terms that will vanish upon the $X$ measurement. We see that while the errors are modified by $\vec{y}$, the correlation between the two sides of $\rho_Z$ that H-VEC utilises remains intact. We may then proceed to apply a stabiliser measurement that produces the syndrome $\vec{k} \in \mathbb{E}_{\mathrm{cor}}$, which leads to
\begin{align}\label{eq: rho_k1_prime}
\begin{split}
    \rho'_{\vec{k},1} =& \frac{1}{P'_{\vec{k}}} X^{\vec{k}}\Pi_{Z}X^{\vec{k}}\rho'_{\text{full},1}X^{\vec{k}}\Pi_{Z}X^{\vec{k}} \\
    =& \frac{1}{2P'_{\vec{k}}} \bigg[...+ \ketbra{0}{1} \otimes \sum_{\vec{x},\vec{y},\vec{z} \in \mathbb{E}_{\mathrm{cor}}} p_{\vec{x},\vec{y},\vec{z}}\\
   &(-1)^{\vec{x} \cdot \vec{z}} \delta_{\vec{x}\oplus \vec{y},\vec{k}}\delta_{\vec{z} \oplus \vec{y},\vec{k}} X^{\vec{k}} Z^{\vec{z}\oplus \vec{y}} \rho_Z Z^{\vec{x}\oplus \vec{y}}X^{\vec{k}} + c.c. \bigg]\\
   =& \frac{1}{2P'_{\vec{k}}} \bigg[...+ \ketbra{0}{1} \otimes \sum_{\vec{y} \in \mathbb{E}_{\mathrm{cor}}^{(\vec{k})}} p_{\vec{y}\oplus \vec{k},\vec{y},\vec{y}\oplus \vec{k}}\\
   &(-1)^{|\vec{k}|}(-1)^{|\vec{y}|}X^{\vec{k}} Z^{\vec{k}} \rho_Z Z^{\vec{k}}X^{\vec{k}} + c.c. \bigg],
\end{split}
\end{align}
where $P'_{\vec{k}}$ is the probability of measuring syndrome $\vec{k}$ and we defined the set of errors $\mathbb{E}_{\mathrm{cor}}^{(\vec{k})} = \{\vec{y} \in \mathbb{E}_{\mathrm{cor}} : \vec{y}\oplus\vec{k} \in \mathbb{E}_{\mathrm{cor}}\}$. Now, being unaware of the $Y$ errors during stabiliser measurement, we apply the corrections $(-1)^{|\vec{k}|}$ and $Y^{\vec{k}}$ and obtain
\begin{align*}
\begin{split}
    \rho'_{\vec{k},2} =& (-1)^{|\vec{k}|}Y^{\vec{k}}\rho'_{\vec{k},1}Y^{\vec{k}} \\
   =& \frac{1}{2P'_{\vec{k}}} \bigg[...+ \ketbra{0}{1} \otimes \sum_{\vec{y} \in \mathbb{E}_{\mathrm{cor}}^{(\vec{k})}} p_{\vec{y}\oplus \vec{k},\vec{y},\vec{y}\oplus \vec{k}}(-1)^{|\vec{y}|} \rho_Z + c.c. \bigg].
\end{split}
\end{align*}
Just as in \cref{eq: rho_k2}, we see that both $X$- and $Z$-type errors are corrected. While there is an uncorrected phase of $(-1)^{\vec{y}}$, this will be cancelled in post-processing, albeit at the cost of additional sampling overhead. It is also worth noting that no uncorrectable errors have gained a finite error probability.

The mixture of all possible $\vec{k}$ is
\begin{align*}
    &\rho'_{\mathrm{full},2}= \sum_{\vec{k}\in \mathbb{E}_{\mathrm{cor}}} P'_{\vec{k}} \rho'_{\vec{k},2}\\
    =& \frac{1}{2} \bigg[...+ \ketbra{0}{1} \otimes \sum_{\vec{k}\in \mathbb{E}_{\mathrm{cor}}}\sum_{\vec{y} \in \mathbb{E}_{\mathrm{cor}}^{(\vec{k})}} p_{\vec{y}\oplus \vec{k},\vec{y},\vec{y}\oplus \vec{k}}(-1)^{|\vec{y}|} \rho_Z + c.c. \bigg].
\end{align*}
Thus, the numerator of our desired ratio corresponds to
\begin{align*}
    &\Tr((X \otimes O) \rho'_{\mathrm{full},2})\\
    =& \bigg{(}\sum_{\vec{k}\in \mathbb{E}_{\mathrm{cor}}}\sum_{\vec{y} \in \mathbb{E}_{\mathrm{cor}}^{(\vec{k})}} p_{\vec{y}\oplus \vec{k},\vec{y},\vec{y}\oplus \vec{k}}(-1)^{|\vec{y}|}\bigg{)}\Tr(O\rho_Z),
\end{align*}
such that we exactly recover $\Tr(O\rho_Z)$ upon dividing by the case where $O$ is replaced by $I$.

In practice, these biased errors that occur during stabiliser measurement can be effectively suppressed by repeating the checks as in usual QEC. Since we considered the biased error channel to occur right before the subsequent parity check, the above result guarantees (under a few assumptions) that repeating checks without repeating the C-H layers remains effective. Suppose that we now want to repeat the pairs of C-H layers to limit the amount physical errors that need to be corrected per round. In this case, a biased error channel that occurs right after a check can still be corrected in the subsequent round, since that would modify \cref{eq: rho_full1} into

\begin{align*}
\begin{split}
    \rho_{\text{full},1}''=& \frac{1}{2} \bigg[...+ \ketbra{0}{1} \otimes \sum_{\vec{x},\vec{y},\vec{z} \in \mathbb{E}_{\mathrm{cor}}} p_{\vec{x},\vec{y},\vec{z}}\\
   &(-1)^{\vec{x} \cdot \vec{z}} X^{\vec{x}} Z^{\vec{z}} Y^{\vec{y}} \rho_Z Y^{\vec{y}} Z^{\vec{x}}X^{\vec{z}} + c.c. \bigg],
\end{split}
\end{align*}
which is equivalent to \cref{eq: rho_full_prime}.

\section{Details of General Framework}\label{sec:derive_gen_framework}
\subsection{Derivation}
From \cref{eq: assumption 1} one can show that the output immediately after the controlled-$U^\dagger$ (C-$U^\dagger$) gate in \cref{fig: general circ} is
\begin{align*}
   \frac{1}{2}\sum_{i,j}  p_{i,j} \bigg[ \ketbra{0} \otimes E_i F_j\rho F_j^{\dagger} E_i^\dagger + \beta_{ij}^*&\ketbra{0}{1} \otimes E_i F_j \rho F_i^\dagger E_j^{\dagger} \\
   + \beta_{ij}  \ketbra{1}{0} \otimes E_j F_i \rho F_j^{\dagger} E_i^\dagger
   + |\beta_{ij}|^2&\ketbra{1} \otimes E_j F_i \rho F_i^\dagger E_j^{\dagger}
   \bigg]
\end{align*}
for any input state $\rho$. After post-processing with an $X$ measurement of the control qubit, we have the effective output
\begin{align}\label{eq: general virtual state}
    \frac{1}{2}\sum_{i,j} p_{i,j} \bigg[ \beta_{ij}^*E_i F_j \rho F_i^\dagger E_j^{\dagger}  + \beta_{ij} E_j F_i \rho F_j^{\dagger} E_i^\dagger\bigg].
\end{align}

Suppose the input state $\rho$ is in the code space $\Pi_E$ and that a stabiliser measurement (denoted $S_E$ in \cref{fig: general circ}) performed immediately after the C-$U^\dagger$ gate produces the syndrome $k$, indicating that the output state is in the syndrome subspace $E_k\Pi_E E_k^\dagger$. Then, up to scalar factors, the second term of \cref{eq: general virtual state} is projected as
\begin{align}\label{eq: general projection}
\begin{split}
    & (E_k\Pi_E E_k^\dagger) E_jF_i\rho F_j^\dagger E_i^\dagger (E_k\Pi_E E_k^\dagger)\\
    =& (E_k\Pi_E E_k^\dagger) E_jF_i(\Pi_E\rho \Pi_E)F_j^\dagger E_i^\dagger (E_k\Pi_E E_k^\dagger)\\
    =& E_k(\Pi_E E_k^\dagger E_j\Pi_E)F_i \Pi_E\rho \Pi_E F_j^\dagger (\Pi_E E_i^\dagger E_k\Pi_E)E_k^\dagger\\
    =& \delta_{ik}\delta_{jk} \lambda_i \lambda_j E_kF_i\rho F_j^\dagger E_k^\dagger,
\end{split}
\end{align}
where we have used \cref{eqn:kl_cond,eqn:undetectable_cond_1}. The first term is similarly projected, such that the effective output state is given by
\begin{equation}
     p_{k,k} \lambda_k^2\bigg[ \mathrm{Re}[\beta_{kk}]E_kF_k\rho F_k^\dagger E_k^\dagger\bigg].
\end{equation}

\subsection{Possible Ways to Find the Transformation Unitary}
One way for \cref{eq: assumption 1} to be true is by having $\{F_i\}$ and $\{E_i\}$ connected via the conjugation of some unitary $U^\dagger$ (note that both sides in the formula below are $U^\dagger$):
\begin{equation}\label{eqn:e_and_f_relation}
    F_i = U^{\dagger} E_i U^{\dagger} \quad \forall i,
\end{equation}
and also if $\{E_i\}$ and $\{F_j\}$ further satisfy
\begin{align}\label{eqn:undetectable_cond_2}
    F_iE_j = \beta_{ij} E_j F_i \quad \forall i, j
\end{align}
for some complex number $\beta_{ij}$. Note that this equation trivially holds for all Pauli errors. Combining \cref{eqn:e_and_f_relation} and \cref{eqn:undetectable_cond_2}, we will arrive at \cref{eq: assumption 1}.

Even if \cref{eqn:e_and_f_relation} or \cref{eqn:undetectable_cond_2} are not true, \cref{eq: assumption 1} can still be satisfied as can be seen in this next example. Although intuitively the use of Hadamard gates in H-VEC can be viewed as a way to combine bit-flip and phase-flip codes, within the general framework, this choice of $U$ is motivated by the fact that it transforms the error components as in \cref{eq: assumption 1}. In fact, the latter can also be achieved using $\sqrt{Y}$ gates, despite not satisfying \cref{eqn:e_and_f_relation} as $\sqrt{Y}^{\otimes n} X^{\vec{x}} \sqrt{Y}^{\otimes n} \neq \sqrt{Y}^{\otimes n} X^{\vec{x}} (\sqrt{Y}^{\otimes n})^\dagger = Z^{\vec{x}}$. We show this by verifying that \cref{eq: assumption 1} still holds:
\begin{align}
    \begin{split}
        &(\sqrt{Y}^{\otimes n})^\dagger X^{\vec{x}} Z^{\vec{z}} \sqrt{Y}^{\otimes n}\\
        =& [(\sqrt{Y}^{\otimes n})^\dagger X^{\vec{x}}\sqrt{Y}^{\otimes n}][(\sqrt{Y}^{\otimes n})^\dagger Z^{\vec{z}} \sqrt{Y}^{\otimes n}]\\
        =& (Z^{\vec{x}})^\dagger [(-1)^{|\vec{z}|}X^{\vec{z}}]^\dagger\\
        =& (-1)^{|\vec{z}|}(-1)^{\vec{x}\cdot \vec{z}}X^{\vec{z}}Z^{\vec{x}}
    \end{split}
\end{align}
This implies that, with all other components in \cref{fig: vqc} unchanged, replacing $H^{\otimes n}$ with $\sqrt{Y}^{\otimes n}$ achieves the same error suppression using the same amount of resources as H-VEC. Since the phase factors cancel upon projection, we do not need to apply the additional phase correction in this case. The same holds for the variants of H-VEC in general Pauli bases as discussed in \cref{sec: bias}, where $H_{\sigma' \sigma''}^{\otimes n}$ can be replaced by $\sqrt{\sigma}^{\otimes n}$.

\subsection{Mapping to Existing Protocols}\label{subsec: mapping}

\begin{figure}
\includegraphics[width=0.38\textwidth]{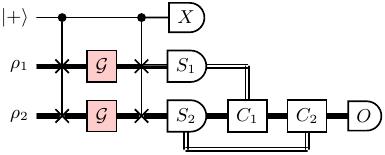}
\caption{Quantum circuit involved in SWAP-VEC, proposed in Ref.~\cite{liu2025virtual}, which enables protecting a logical register using two QEC codes together.}\label{fig: vec}
\end{figure}

Let us now reformulate the two known protocols within the general VEC framework.

\paragraph{H-VEC.} As presented in \cref{sec: vqc}, $\rho$ corresponds to some code state and $\Pi_E$ is the code space projector of a bit-flip classical code. The noise channel $\mathcal{N}$ is a Pauli error channel as in \cref{eq: pauli}. The set $\{E_i\}$ ($\{F_i\}$) then corresponds to all bit-flip (phase-flip) errors with support that lies in $\mathbb{E}_{\mathrm{cor}}$. By choosing the indices of $\{E_i\}$ and $\{F_i\}$ to correspond to the binary vectors that define their supports, we have $H^{\otimes n}E_iF_jH^{\otimes n} = F_iE_j = (-1)^{i\cdot j}E_jF_i$ for all $i, j$, which means \cref{eq: assumption 1} is satisfied for $U = H^{\otimes n}$ and $\beta_{i,j} = (-1)^{i\cdot j}$.

\paragraph{SWAP-VEC.} Ref.~\cite{liu2025virtual} considers two quantum registers, each encoded in some code state $\rho_l$ of a code defined by the code space projector $\Pi_l$ for $l\in \{1,2\}$. In the notation of our framework, we thus have $\rho = \rho_1 \otimes \rho_2$ and $\Pi_E = \Pi_1\otimes \Pi_2$. The set of correctable errors for code $\Pi_1$ of the first register is denoted as $\{G_{i}^{(1)}\}$ and similarly that of the second register is $\{G_{i}^{(2)}\}$. Suppose we choose $\Pi_1$, $\Pi_2$ and also their set of non-degenerate correctable errors such that all $G_{i}^{(2)}$ are not detectable by $\Pi_1$ and similarly, all $G_{i}^{(1)}$ are not detectable by $\Pi_2$. Then, by focusing on Pauli errors, we can show that \cref{eq: assumption 1} is satisfied by defining $E_i = G_{i_1}^{(1)} \otimes G_{i_2}^{(2)}$ and $F_i = G_{i_2}^{(2)} \otimes G_{i_1}^{(1)}$, where $i_1$ and $i_2$ need not be equal in general, and with $U$ being the SWAP operator between the two registers. Hence, the VEC protocol can be carried out using the circuit in \cref{fig: vec}. As mentioned in Ref.~\cite{liu2025virtual}, one practical choice is to have a bit-flip code and a phase-flip code in the two registers, respectively.

\section{Devices with Biased Noise \label{sec: bias}}

As discussed in \cref{sec:performance}, in many aspects, the strong error suppression and the sampling overhead of our protocol can be viewed as the result of effectively post-selecting for pure $Y$ errors, and the latter Pauli basis can be changed to $X$ or $Z$ as well by making appropriate modifications. Since the sampling overhead depends strongly on the probability of errors towards the specific noise-bias (see \cref{eqn:vec_example_overhead}), H-VEC is particularly suited to biased-noise systems.

In this case, as shown in \cref{app: repeat}, it is also much easier to deal with the errors that occur in the stabiliser checks as we go beyond the code-capacity error model. Importantly, under biased noise, we can repeat H-VEC and stabiliser checks without compromising much of its performance. Even though the noise in the stabiliser checks are outside the pair of C-H layers and thus not acted on by H-VEC, the dominant $Y$ noise can still be detected and corrected by the next round of checks without needing additional actions. Although weaker due to the noise bias, $X$ and $Z$ noise in each check is also left unfiltered by H-VEC, and errors in these channels will accumulate with repeated rounds of checks. However, due to the simple nature of the checks required for classical codes, such check noise accumulation will be relatively slow, and should enable some interesting application scenarios with repeated checks for systems with significantly biased noise.

Given a quantum device with bias towards $\sigma$-type errors for $\sigma \in \{X,Y,Z\}$, H-VEC can be easily adapted to suit the device without increasing the resource requirements. Indeed, minor adjustments lead to compatibility with, e.g., trapped-ion platforms with a more typical bias towards $Z$-type errors~\cite{langer2005long,sepiol2019probing,seis2023balancing}. The modification consists of two parts, described below.

Firstly, we adjust the Pauli basis of the code space projector. In \cref{sec: vqc}, we have chosen the input code stabiliser basis to be $Z$ with the corresponding code space projector denoted as $\Pi_Z$. For a more general biased noise basis $\sigma$, we need to choose the stabiliser basis of the input code to be $\sigma' \neq \sigma$, which is different from the noise basis. Secondly, we replace the Hadamard in C-H with
\begin{equation}
    H_{\sigma' \sigma''} = \frac{1}{\sqrt{2}}(\sigma' + \sigma''),
\end{equation}
where $\sigma''$ is the remaining Pauli basis that is different from both the noise basis $\sigma$ and the input code basis $\sigma'$.

It is easy to see that for the $\sigma = Y$ case, we recover $\Pi_Z$ or $\Pi_X$ for the input code state, and $H_{X Z} = H$ for the C-H gates. We remark that the implementation described here is not unique. In fact, we show in \cref{sec:derive_gen_framework} that the $\sigma$ version of H-VEC can also be implemented using controlled-$\sqrt{\sigma}$ gates instead.

\section{Connectivity-flexible Unit-depth C-H Layers via
Multiple Unentangled Control qubits}\label{app: transversal}

\begin{figure}[htbp]
\includegraphics[width=0.48\textwidth]{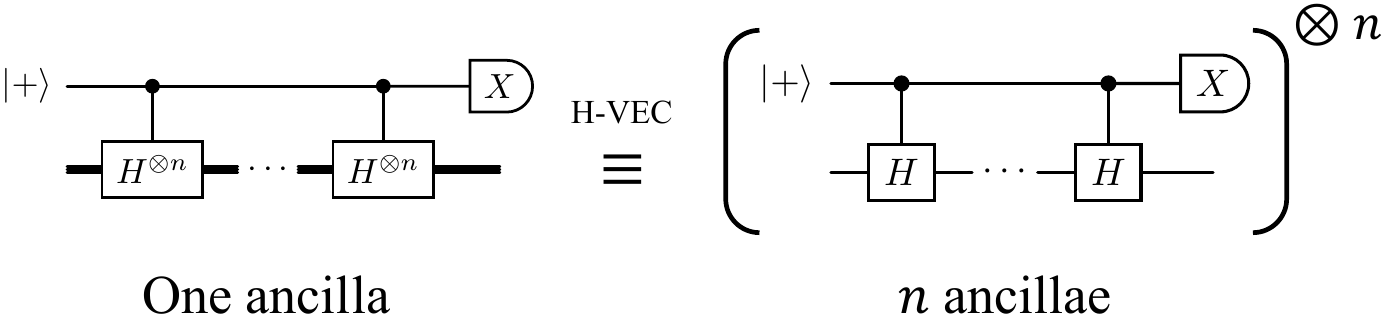}
\caption{The circuit in which C-H gates are applied on the whole register using one control qubit can be replaced by a circuit consisting of one control qubit per physical qubit, without altering the performance of H-VEC. The original control qubit measurement outcome simply corresponds to the product between the outcomes of all $n$ control qubits.}\label{fig: CH_decomp}
\end{figure}

As discussed in \cref{subsec: limits}, the single control qubit of H-VEC can be replaced by $n$ unentangled ones to parallelise the C-H layers, as shown in \cref{fig: CH_decomp}. In this case, each physical qubit of the $n$-qubit logical register is paired with a single control qubit. For each pair, two C-H gates are applied to the qubit in the logical register conditioned on the control qubit state, such that each of the two C-H gate layers can be performed in a single parallelised step. The logical register is still protected from a Pauli error channel sandwiched by the two C-H gate layers upon the same stabiliser measurement and corrections. Correspondingly, the measurement at the end will be changed from $X \otimes O$ to $X^{\otimes n} \otimes O$. In the following, we show why this replacement can be performed.

Using $\ketbra{+}^{\otimes n} = \frac{1}{2^{n}}\sum_{\vec{a},\vec{b} \in \{0,1\}^n} \ketbra{\vec{a}}{\vec{b}}$, the output of the transversal CH variants of \cref{fig: vqc} before the $X$ measurement is
\begin{align*}
   \rho_{\text{full},1} &= \frac{1}{2^{n}}\sum_{\vec{x}, \vec{{z}} \in \mathbb{E}_{\mathrm{cor}}}p_{\vec{x}, \vec{z}}  \sum_{\vec{a},\vec{b} \in \{0,1\}^n} \ketbra{\vec{a}}{\vec{b}} \\
   &\quad \otimes \left(H^{\vec{a}} X^{\vec{x}}  Z^{\vec{z}}H^{\vec{a}}\right) \rho_Z \left(H^{\vec{b}}Z^{\vec{z}} X^{\vec{x}}H^{\vec{b}}\right)
\end{align*}
Note that $\ketbra{\vec{a}}{\vec{b}} = \bigotimes_{i = 1}^n \ketbra{a_i}{b_i}$ where $a_i$ and $b_i$ are the $i$th digits of $\vec{a}$ and $\vec{b}$. Now the measurement of $X^{\otimes n}$ will remove any terms of with overlapping digits between $\vec{a}$ and $\vec{b}$, i.e. any term with $a_i = b_i$ for any $i$ since $\Tr(\ketbra{a_i}{b_i} X) = 0$ when $a_i = b_i$. So after performing $X^{\otimes n}$ measurement on the control qubits, the remaining terms have $\vec{b} = \vec{1}\oplus\vec{a}$ for all $\vec{a}$ and $\vec{b}$, and the effective output states becomes:
\begin{align*}
   \rho_{\text{coh}} &\propto \frac{1}{2^{n}}\sum_{\vec{x}, \vec{{z}} \in \mathbb{E}_{\mathrm{cor}}}p_{\vec{x}, \vec{z}}   \\
   &\quad \sum_{\vec{a}\in \{0,1\}^n} \left(H^{\vec{a}} X^{\vec{x}}  Z^{\vec{z}}H^{\vec{a}}\right) \rho_Z H^{\otimes n}\left(H^{\vec{a}}Z^{\vec{z}} X^{\vec{x}} H^{\vec{a}}\right)H^{\otimes n}
\end{align*}
where we have use $H^{\vec{1} \oplus \vec{a}} = H^{\otimes n} H^{\vec{a}}$.

We can see that this is in a similar form as the single control-qubit case with $X^{\vec{x}}  Z^{\vec{z}}$ replaced by
\begin{align*}
    H^{\vec{a}} X^{\vec{x}}  Z^{\vec{z}}H^{\vec{a}}  = (-1)^{\alpha} X^{\vec{u}}  Z^{\vec{v}}
\end{align*}
Here $\vec{u}$ is simply takes the entry of $\vec{z}$ at the support of $\vec{a}$ and takes the entry of $\vec{x}$ at the support of $\neg \vec{a}$. Similarly for the others. Thus we have:
\begin{align*}
    \vec{u} &= (\neg\vec{a})\land\vec{x} \oplus \vec{a}\land\vec{z}\\
    \vec{v}&=(\neg\vec{a})\land\vec{z}\oplus\vec{a}\land\vec{x}\\
    \alpha &= \vec{a} \cdot (\vec{u} \land \vec{v})
\end{align*}
Since $H^{\vec{a}}$ self-inverse, we also have:
\begin{align*}
      H^{\vec{a}}X^{\vec{u}}  Z^{\vec{v}}H^{\vec{a}} = (-1)^{\alpha} X^{\vec{x}}  Z^{\vec{z}}
\end{align*}
\begin{equation}
    \begin{split}
        \vec{x} &= (\neg\vec{a})\land\vec{u} \oplus \vec{a}\land\vec{v} \label{eqn:x_in_u}\\
    \vec{z}&=(\neg\vec{a})\land\vec{v}\oplus\vec{a}\land\vec{u}\\
    \alpha &= \vec{a} \cdot (\vec{u} \land \vec{v})
    \end{split}
\end{equation}

Note that $(\vec{x}, \vec{z}) \mapsto (\vec{u}, \vec{v})$ is a bijective map for a given $\vec{a}$ with $(\vec{x}, \vec{z}) \in \mathbb{E}_{\mathrm{cor}}^2$ and $(\vec{u}, \vec{v}) \in \widetilde{\mathbb{E}}_{\mathrm{cor}}$. Also note that $\widetilde{\mathbb{E}}_{\mathrm{cor}}$ is not necessarily decomposable into tensor product of two sets like $\mathbb{E}_{\mathrm{cor}}^2$.

For example, for a three-qubit system, if we have $\mathbb{E}_{\mathrm{cor}} = \{[0,0,0], [0,0,1], [0,1,0], [1,0,0]\}$, i.e. we can correct all weight $1$ error. And we have $\vec{a} = [1,0,0]$, i.e. we are exchanging the $X$ and $Z$ on the first qubit. We will replace the following entries in $\mathbb{E}_{\mathrm{cor}}^2$:
\begin{align*}
    ([0,1,0],[1,0,0]) &\mapsto ([1,1,0],[0,0,0])\\
    ([1,0,0],[0,1,0]) &\mapsto ([0,0,0],[1,1,0])\\
    ([0,0,1],[1,0,0]) &\mapsto ([1,0,1],[0,0,0])\\
    ([1,0,0],[0,0,1]) &\mapsto ([0,0,0],[1,0,1])
\end{align*}
and the resultant set $\widetilde{\mathbb{E}}_{\mathrm{cor}}$ cannot be written as a tensor product of two sets since $[1,1,0]$ (or $[1,0,1]$) will only pair with $[0,0,0]$ and nothing else in $\widetilde{\mathbb{E}}_{\mathrm{cor}}$.

We will further define 
\begin{align*}
    q_{\vec{u}, \vec{v}} = p_{\vec{x}, \vec{z}}
\end{align*}
with $\vec{x}$ and $\vec{z}$ on the L.H.S given by \cref{eqn:x_in_u}.

Hence, we have

\begin{align*}
   &\rho_{\text{coh}} 
   = \frac{1}{2^{n}}\\
   &\ \ \times \sum_{\vec{a}\in \{0,1\}^n} \sum_{(\vec{u}, \vec{v}) \in \widetilde{\mathbb{E}}_{\mathrm{cor}}} q_{\vec{u}, \vec{v}} \left(X^{\vec{u}}  Z^{\vec{v}}\right) \rho_Z H^{\otimes n}\left(Z^{\vec{v}} X^{\vec{u}}\right)H^{\otimes n}\\
   & = \frac{1}{2^{n}}\sum_{\vec{a}\in \{0,1\}^n} \sum_{(\vec{u}, \vec{v}) \in \widetilde{\mathbb{E}}_{\mathrm{cor}}} q_{\vec{u}, \vec{v}} (-1)^{\vec{u} \cdot \vec{v}} \left(X^{\vec{u}}  Z^{\vec{v}}\right) \rho_Z \left(Z^{\vec{u}} X^{\vec{v}}\right)
\end{align*}

Since both $\vec{x}$ and $\vec{z}$ are correctable, we know that $\vec{x} \oplus \vec{z}$ will not form a logical of the classical code. Using the fact that $\vec{u} \oplus \vec{v} = \vec{x} \oplus \vec{z}$, we know that $\vec{u} \oplus \vec{v}$ will not be a logical as well. We are considering a non-degenerate code, thus only errors differ by a logical can have the same syndrome. Since we know that  $\vec{u}$ and $\vec{v}$ cannot be differed by a logical, the only way $\vec{u}$ and $\vec{v}$ to have the same syndrome is to have $\vec{u} = \vec{v}$. Hence, once we perform syndrome measurement and trying to project both side of $X^{\vec{u}}$ and $X^{\vec{v}}$ into the same syndrome subspace, only terms with  $\vec{u} = \vec{v}$ will remain, just the same as before. 

What are the remaining terms then? Applying the inverse mapping from $(\vec{u}, \vec{v})$ to $(\vec{x}, \vec{z})$, we found that all terms with $\vec{u} = \vec{v}$ not change under this map, i.e. they will map to $(\vec{x}, \vec{z})$ that has $\vec{x} = \vec{z} = \vec{u} = \vec{v}$. So this is just all of our previous terms in the single-control version! Hence, all of these terms will produce the same syndrome as before and we can correct using the same decoders and phase correction. The same set of remaining term also means the same error suppression performance and sampling overhead.

Similar argument also applies to the SWAP-based case, but now we have two correctable sets $\mathbb{E}_{\mathrm{cor}}^X$ and $\mathbb{E}_{\mathrm{cor}}^Z$ for the two code register, and we also have the error probability of the two Pauli error channel on the two register be $p^{(1)}_{\vec{x}, \vec{z}}$  and $p^{(2)}_{\vec{r}, \vec{s}}$ respectively:
\begin{align*}
    \rho_{\text{coh}} &\propto \frac{1}{2^{n}}\sum_{\vec{x}, \vec{r} \in \mathbb{E}_{\mathrm{cor}}^{X}}\sum_{\vec{z}, \vec{s} \in \mathbb{E}_{\mathrm{cor}}^{Z}}p^{(1)}_{\vec{x}, \vec{z}}p^{(2)}_{\vec{r}, \vec{s}}\sum_{\vec{a}\in \{0,1\}^n}   \\
   &\quad 
    \text{SW}^{\vec{a}}\left(X^{\vec{x}}  Z^{\vec{z}}\right) \otimes \left(X^{\vec{r}}  Z^{\vec{s}}\right)\text{SW}^{\vec{a}} \\
    &\quad\times \rho \text{SW}^{\vec{a}}\left(X^{\vec{r}}  Z^{\vec{s}}\right) \otimes \left(X^{\vec{x}}  Z^{\vec{z}}\right)\text{SW}^{\vec{a}}
\end{align*}
We will use
\begin{align*}
    \text{SW}^{\vec{a}}\left(X^{\vec{x}}  Z^{\vec{z}}\right) \otimes \left(X^{\vec{r}}  Z^{\vec{s}}\right)\text{SW}^{\vec{a}} = \left(X^{\vec{u}}  Z^{\vec{v}}\right) \otimes \left(X^{\vec{w}}  Z^{\vec{j}}\right)
\end{align*}
where $\vec{u}$ is simply takes the entry of $\vec{r}$ at the support of $\vec{a}$ and takes the entry of $\vec{x}$ at the support of $\neg \vec{a}$. Hence, we have:
\begin{align*}
    \vec{u} &= (\neg\vec{a})\land\vec{x} \oplus \vec{a}\land\vec{r}\\
    \vec{w} &= (\neg\vec{a})\land\vec{r} \oplus \vec{a}\land\vec{x}\\
    \vec{v}&=(\neg\vec{a})\land\vec{z}\oplus\vec{a}\land\vec{s}\\
    \vec{j}&=(\neg\vec{a})\land\vec{s}\oplus\vec{a}\land\vec{z}
\end{align*}
Focus on the $X$ errors and thus $\vec{u}$ and $\vec{w}$, following the same arguments before, we have $\vec{u} + \vec{w} = \vec{x} + \vec{r}$ and thus $\vec{u} + \vec{w}$ will not be a logical. Thus, by projecting using $X^{\vec{k}} \Pi_Z X^{\vec{k}}$ on the first register, only the terms with both $X^{\vec{u}}$ and $X^{\vec{w}}$ being correctable will remain on the first register. The second register will also be projected in the the same subspace since its noise is entangled with the first register.

\section{Protecting Resource States}\label{app: magic}

We have seen the potential of H-VEC in correcting quantum noise on an arbitrary input state encoded using a classical code, which can enable resource-efficient long-range lattice surgery. Here, we present a different class of application. In many important scenarios, we are interested in protecting specific states, which may bring about further simplification of the H-VEC circuit. In these cases, we often want to protect a state that is encoded in a quantum code rather than a classical code, and thus we are unable to reap the full benefit of H-VEC. Nevertheless, as we will show in the entanglement purification example below, there are scenarios in which H-VEC can still achieve some resource savings.

\begin{figure}
\centering
    \includegraphics[width=0.48\textwidth]{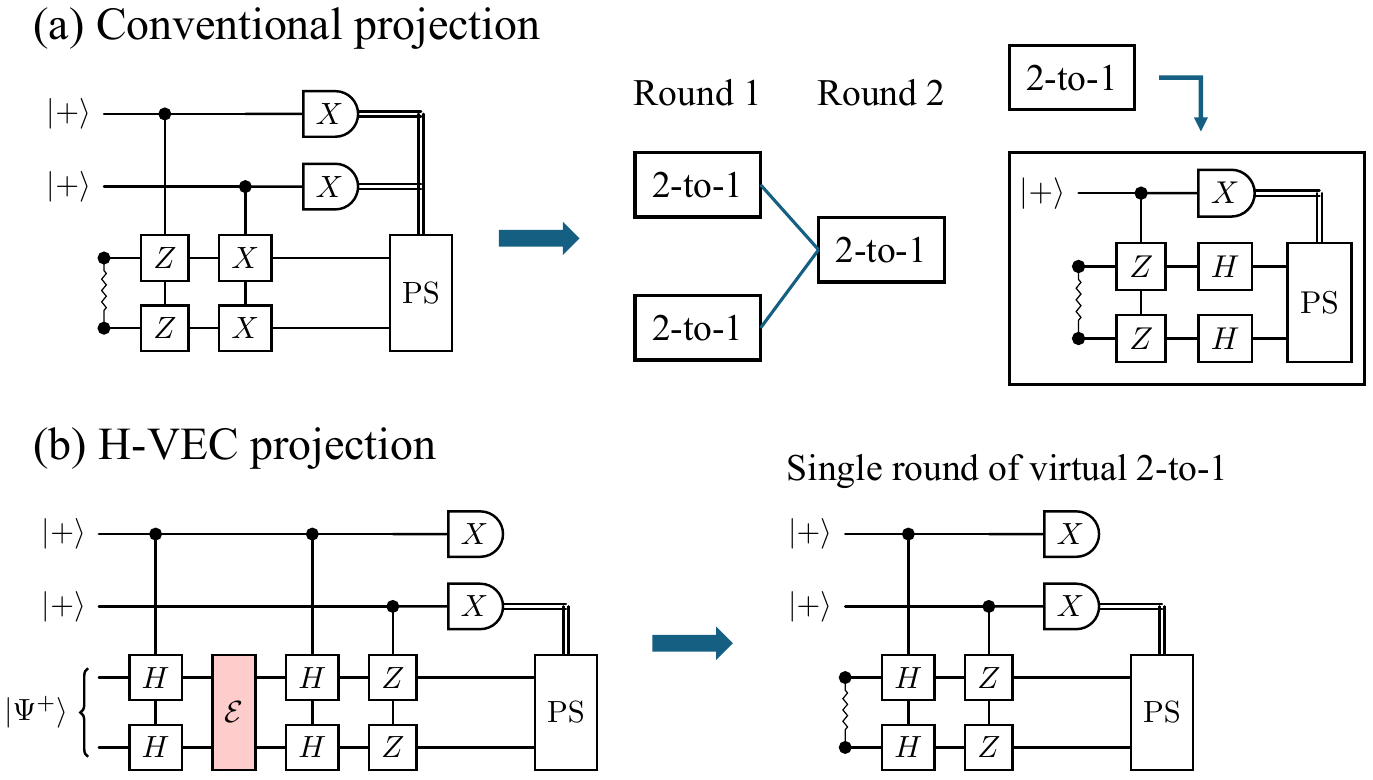}
    \caption{EPPs viewed in terms of stabiliser projections performed in the (a)~conventional and (b)~H-VEC approach. Each vertical wavy line represents a noisy Bell pair $\mathcal{E}\big{[}\ketbra{\Psi^+}\big{]}$ established between two qubits, and ``PS'' denotes post-selection. (a)~The basic 2-to-1 stabiliser EPP~\cite{shi2025stabilizer} sequentially measures the $Z\otimes Z$ and $X \otimes X$ stabilisers of the Bell state by interleaving Hadamard gates between two rounds. (b)~When applying H-VEC to $\ket{\Psi^+}$, the first C-H layer can be removed so it can be applied directly to the noisy state rather than the noise channel. Only a single round is required in this case, leading also to a reduction in the number of noisy input Bell states.}\label{fig: vepp}
\end{figure}

\subsection{Application to Entanglement Purification}

In entanglement purification, the objective is to prepare high-fidelity Bell states $\ket{\Psi^+} = (1/\sqrt{2})(\ket{00}+\ket{11})$ from a larger number of noisy ones via an entanglement purification protocol (EPP), which is a crucial ingredient in modular quantum architectures and quantum communication. The Bell state $\ket{\Psi^+}$ is stabilised by $Z\otimes Z$ and $X \otimes X$, and many EPPs can be viewed as applying multiple rounds of fanned-out stabiliser projections using noisy Bell pairs as control qubits~\cite{dur2007entanglement,bennett1996purification,deutsch1996quantum}. Recently, Ref.~\cite{shi2025stabilizer} viewed basic recurrence-type EPPs in this stabiliser perspective, and described a general construction that consists of two rounds as illustrated in \cref{fig: vepp}~(a). In the first round of a 2-to-1 protocol, a $Z\otimes Z$ check is applied to a noisy Bell pair using another noisy Bell pair that corresponds to a fanned-out control qubit, obtaining an intermediate Bell pair. Two other noisy Bell states are used to obtain another intermediate Bell pair, before moving on to the second round, which effectively performs an $X\otimes X$ check using these two intermediate Bell pairs and Hadamard gates. Therefore, in total, the protocol consumes $4$ noisy Bell pairs to produce one purified Bell pair over two rounds. 

Now, let us see if we can replace the two rounds of checks with one round of $Z\otimes Z$ check with the help of H-VEC. We start with the circuit depicted on the left side of \cref{fig: vepp}~(b), where the noise on the Bell state is sandwiched between two C-H layers. Such impractical separation between the Bell state and its noise channel can be resolved by noting that $H\otimes H$ applies trivially to $\ket{\Psi^+}$, and thus the first C-H layer can be removed. Only one round of stabiliser check is required after the remaining C-H layer, since pure $X$ and $Z$ noise are filtered out by H-VEC. We only need $3$ Bell states in total: one as the main register, and two others as fanned-out control qubit for C-H gates and $Z\otimes Z$ check, respectively. In fact, the C-H control qubit can be reused as we purify more than one Bell state, as shown in \cref{fig: vepp_2}. Therefore, its cost per individual purified state is negligible, and H-VEC asymptotically reduces the number of input noisy Bell pairs per purified Bell pair from $4$ to $2$.

The virtual EPP introduced here may be investigated further in various directions. For instance, one can study the trade-off between the Bell pair saving and the sampling overhead of H-VEC. It is also worthwhile to consider applications to stabiliser EPPs with higher rates than the 2-to-1 protocol, and examine the extent to which the scheme can surpass the limitations of physical EPPs as reported in other existing virtual EPP proposals~\cite{yuan2024virtual,yamamoto2024virtual}.

Interesting connections can also be made to magic state cultivation~\cite{itogawa2025efficient,chamberland2020very,gidney2024magic}, where a C-H layer is similarly used. When H-VEC is applied to an input magic state in the cultivation context, the first layer of C-H gates again acts trivially and can thus be removed. In this case, we recover exactly the cultivation circuit, but H-VEC employs a different post-processing rule. It will be an interesting direction to study whether applying a different post-processing rule yields any advantage in the context of magic state cultivation.

\subsection{Practical Implementation Beyond Code Capacity}\label{app: vepp_comments}

\begin{figure*}
\centering
    \includegraphics[width=0.95\textwidth]{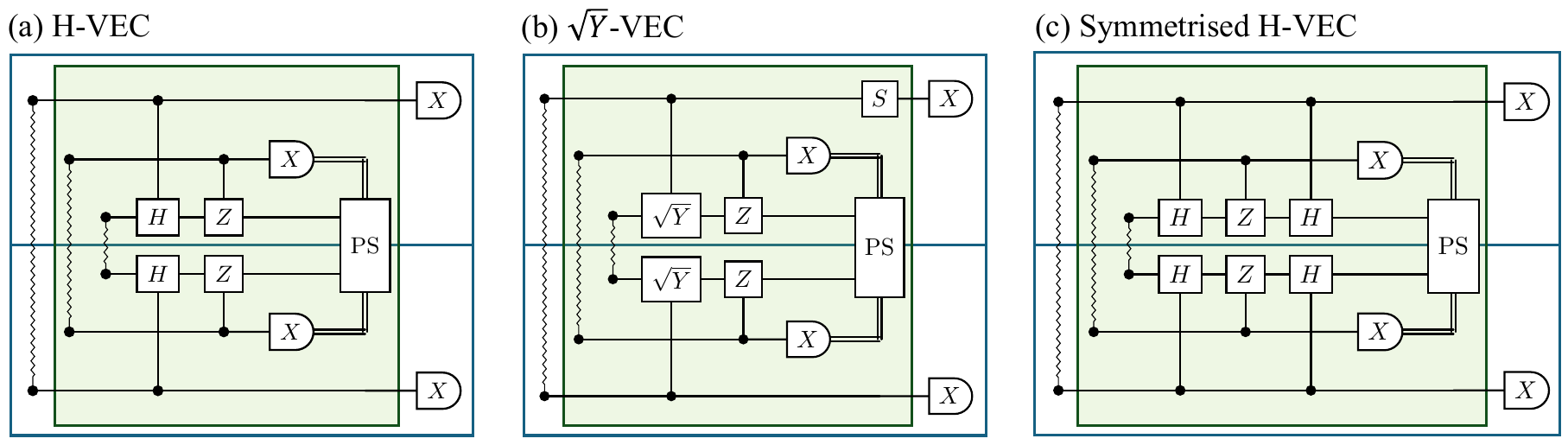}
    \caption{Fanned-out circuits of 2-to-1 virtual EPP based on (a) H-VEC (see \cref{fig: vepp} for an illustration of its stabiliser perspective), (b) $\sqrt{Y}$-VEC, and (c) a symmetrised version of H-VEC where an additional C-H layer is applied after the stabiliser projection. Each vertical wavy line represents a noisy Bell pair $\rho_{\mathrm{noisy}} = \mathcal{E}[\ketbra{\Psi^+}]$ established between two qubits located in different modules (blue boxes). In each case, to purify more than one Bell states, we may reuse the ancilla Bell pair for VEC by sequentially repeating only the components highlighted in green. (a) By fanning out \cite{haner2021distributed} both the control qubits used for stabiliser measurement and for H-VEC, we obtain a circuit that involves fewer quantum hardware resources compared to the 2-to-1 stabiliser EPP in Ref.~\cite{shi2025stabilizer} (replicated in \cref{fig: epp} using $Z$-type stabiliser projections). (b) For Werner state inputs, we can suppress the effects of errors in stabiliser measurements by applying controlled-$\sqrt{Y}$ gates rather than C-H gates for VEC. (c) The performance can be further improved by including an additional layer of C-H gates after the stabiliser projection, which we may apply without altering the ideal performance of H-VEC again due to the invariance of $\ket{\Psi^+}$ under $H\otimes H$.}\label{fig: vepp_2}
\end{figure*}

\begin{figure}
\centering
    \includegraphics[width=0.48\textwidth]{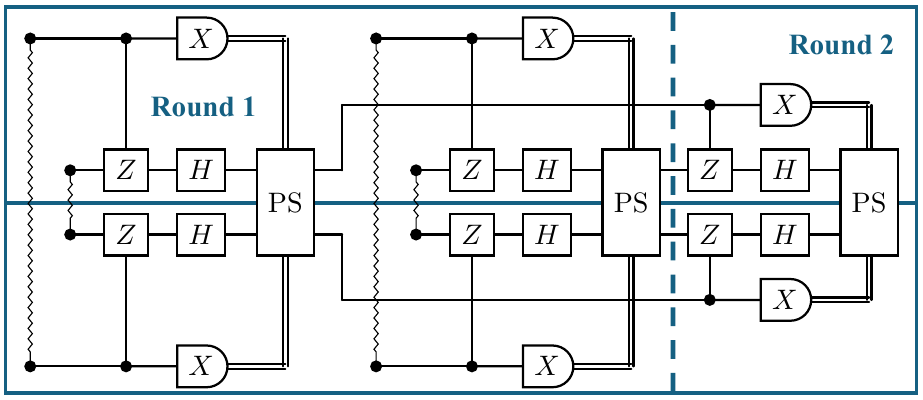}
    \caption{A conventional recurrence-type EPP, which may be viewed as a 2-to-1 protocol that applies stabiliser measurements. In the first round, two pairs of Bell states are used to purify two Bell states. In the second round, the two intermediate purified Bell pairs are consumed to complete the two-round process involving a total of four input noisy Bell pairs. By interleaving Hadamard gates between the two rounds, the noise channel is conjugated such that the circuit is effectively a measurement of both $Z\otimes Z$ and $X\otimes X$ stabilisers.}\label{fig: epp}
\end{figure}

\begin{figure}
\centering
    \includegraphics[width=0.48\textwidth]{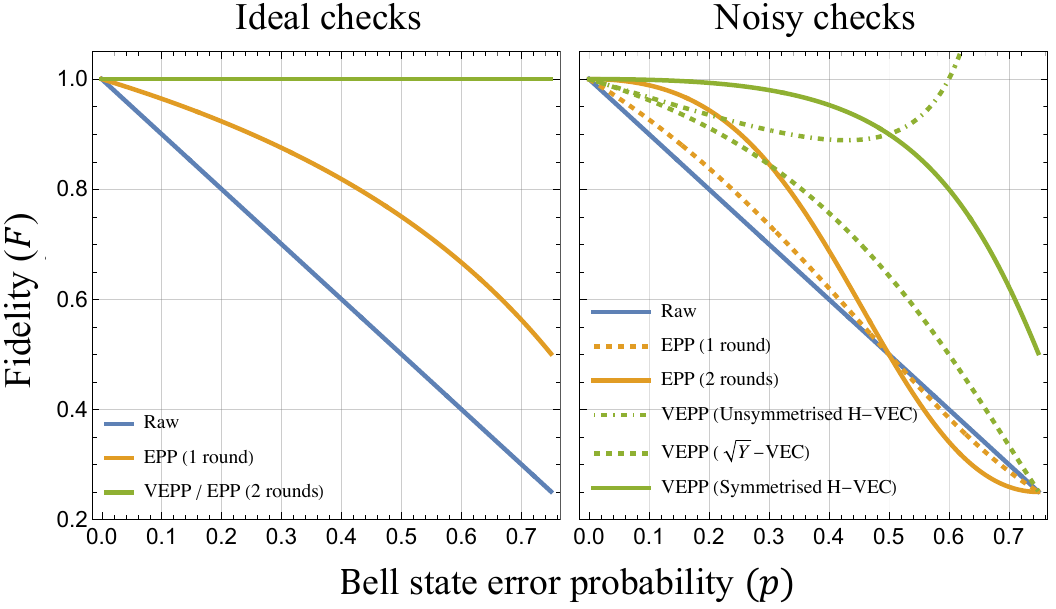}
    \caption{Fidelities between physically or virtually purified Bell states $\rho_{\mathrm{pur}}$ and an ideal Bell state, defined by $F \vcentcolon= \mathrm{Tr}(\ketbra{\Psi^+}\rho_{\mathrm{pur}})/\mathrm{Tr}(\rho_{\mathrm{pur}})$, under a varying error probability $p$ that defines each noisy Bell pair. (left) Comparison between conventional (EPP) and virtual (VEPP) 2-to-1 EPPs when the Bell pair to be purified and the Bell pair for C-H gates are noisy, but the Bell pair for stabiliser check is assumed to be ideal. All variants of VEPP considered achieve unit fidelity under this idealised assumption. The same is achieved if both stabiliser projection rounds of EPP were performed noiselessly. (right) Considering the more practical case where all input Bell pairs are noisy, three variants of VEPP are compared with EPP. We observe that VEPP based on a direct application of H-VEC (green dot-dashed) maintains a high purified fidelity for low error rates, but is less effective than 2 rounds of conventional EPP, and errors during stabiliser projections are detrimental for high error rates. The latter is resolved by applying $\sqrt{Y}$-VEC (green dashed), which does not involved erroneous phases. By symmetrising H-VEC (green solid) via an additional layer of C-H gates after stabiliser projection, we improve the error-suppression robustness of the protocol to surpass that achieved by 2 rounds of EPP.}\label{fig: vepp_sim}
\end{figure}

Now let us consider the more realistic scenario where all input Bell pairs are susceptible to noise. The fanned-out virtual 2-to-1 protocol is shown in \cref{fig: vepp_2}~(a), which may be compared to its physical counterpart shown in \cref{fig: epp}. Although our virtual protocol involves an ancillary Bell state for applying the C-H gates, similarly to Refs.~\cite{tsubouchi2023virtual,yamamoto2024virtual,linErrormitigatedQuantumMetrology2025}, the virtual state is insensitive to a range of noise channels applied to this ancillary Bell state, since the effects are cancelled off in normalisation. In fact, when the noisy ancilla Bell state is only subject to incoherent errors and the Bell state for stabiliser check is assumed ideal, the main virtual state maintains unit fidelity with $\ket{\Psi^+}$, as shown in \cref{fig: vepp_sim}~(left).

To analyse our virtual EPP, we define the fidelity between the purified state $\rho_{\mathrm{pur}}$ and an ideal Bell state as $F \vcentcolon=\mathrm{Tr}(\ketbra{\Psi^+}\rho_{\mathrm{pur}})/\mathrm{Tr}(\rho_{\mathrm{pur}})$, which is bounded between 0 and 1 for physical two-qubit states and is equal to 0.25 for maximally mixed states. In practice, the fidelity is limited by the noisy Bell pair used to apply the stabiliser check, so further adjustments that take such effects into consideration would be crucial. In the following, we will first study how our virtual EPP performs when the stabiliser check is also noisy, and then propose two adjusted circuit designs that may prove useful in practice.

To this end, we assume that noisy Bell pairs are Werner states defined by $(I\otimes\mathcal{D}_p)[\ketbra{\Psi^+}]$ with error probability $p$, noting that we do not lose generality as any two-qubit state can be converted into this form without altering its fidelity by twirling appropriately \cite{bennett1996mixed}. We assume ideal local operations as the fidelity of modular quantum architectures is currently limited by Bell states established between different modules. While we focus on obtaining one purified Bell state, the analysis below also holds for the case where the virtual EPPs are repeated as shown in \cref{fig: vepp_2}, in which case all purified Bell states in the repetition share the same fidelity although the virtually filtered errors and consequently the sampling overhead would accumulate.

First consider applying the virtual EPP as shown in \cref{fig: vepp_2}~(a), assuming that all input noisy Bell pairs are described by the same Werner state determined by $p$. Keeping track of how each error component that applies to otherwise ideal Bell pairs propagate through the circuit, the purified virtual state can be expressed, up to normalisation, as
\begin{align}
\begin{split}
    \rho_{\mathrm{pur}} \propto& \bigg{[}(1-p)^2 I\cdot I + \frac{(1-p)p}{3}Z\cdot Z\\
    &- \frac{p^2}{9}X\cdot X- \frac{p^2}{9}Y\cdot Y\bigg{]}\ketbra{\Psi^+}.
\end{split}
\end{align}
The normalisation factor is thus given by
\begin{align}
\begin{split}
    \mathrm{Tr}(\rho_{\mathrm{pur}}) = (1-p)^2+\frac{(1-p)p}{3}-\frac{2p^2}{9}
\end{split}
\end{align}
and consequently, the fidelity is
\begin{align}\label{eq: F_vepp_1}
\begin{split}
    F_{\mathrm{H}} = \frac{(1-p)^2}{(1-p)^2+(1-p)p/3-2p^2/9}.
\end{split}
\end{align}
The simulated fidelity is plotted as the dot-dashed line in \cref{fig: vepp_sim}~(right), from which we observe that it surpasses unity for large error probabilities. This is due to the denominator approaching $0$ as $p$ increases towards its maximum value of $0.75$, amplifying any errors unaccounted for in the ideal scenario.

As discussed in \cref{sec:derive_gen_framework}, the C-H gates involved in H-VEC can be replaced by controlled-$\sqrt{Y}$ (C-$\sqrt{Y}$) gates with the only difference being that the latter does not involve the erroneous $\pm 1$ phases that require correction. In fact, the minus sign responsible for the vanishing denominator in \cref{eq: F_vepp_1} appears only because of these phases in H-VEC, suggesting that the use of $\sqrt{Y}$-VEC may resolve our aforementioned issue. To do so, we must notice that $\sqrt{Y}\otimes \sqrt{Y}$ only alters $\ket{\Psi^+}$ by a phase as $(\sqrt{Y}\otimes \sqrt{Y}) \ket{\Psi^+} = i\ket{\Psi^+}$, such that we may directly apply $\sqrt{Y}$-VEC to the noisy state similarly to H-VEC but with an additional $S$ gate as shown in \cref{fig: vepp_2}~(b). We then obtain the fidelity
\begin{align}\label{eq: F_vepp_2}
\begin{split}
    F_{\sqrt{Y}} = \frac{(1-p)^2}{(1-p)^2+(1-p)p/3+2p^2/9},
\end{split}
\end{align}
which only differs from \cref{eq: F_vepp_1} by a sign but is now, importantly, bounded between 0 and 1 as desired. As we observe in the dashed line of \cref{fig: vepp_sim}~(right), however, the performance of which is not as high as performing two rounds of conventional EPP in the regime of low error probabilities.

Although the use of our virtual EPP already saves the amount of quantum hardware resources involved, we may consider whether further improvements are possible, particularly in the achievable fidelity. Returning to H-VEC, we may recall that $\ket{\Psi^+}$ is invariant under transformation by $H\otimes H$. This implies that we may apply another layer of C-H gates after the stabiliser projection without disturbing the purified state in the ideal scenario, leading to the symmetrised circuit shown in \cref{fig: vepp_2}~(c). Applying a similar analysis as before, we obtain the fidelity
\begin{align}
\begin{split}
    F_{\mathrm{S-H}} = \frac{(1-p)^2}{(1-p)^2+p^2/9}.
\end{split}
\end{align}
Compared to \cref{eq: F_vepp_2}, we notice that the denominator reduced by $(1-p)p/3 + p^2/9$, which should lead to an increase in the fidelity while ensuring physicality. Indeed, as shown in the solid line of \cref{fig: vepp_sim}~(right), this implementation virtually purifies a Bell state such that the resulting fidelity is not only bounded by 1, but also achieves a strong error suppression power beyond 2 rounds of conventional EPP under errors in all Bell states involved. We note that this circuit can also be viewed as applying the $Z\otimes Z$ stabiliser projection from one side of the noisy density matrix and $X\otimes X$ from the other, for which we expect a close relation to VEC that may be explored in future work.

\section{Details of Numerical Simulations}\label{app: numerics}
\subsection{\cref{fig: threshold-cc}}
We consider a quantum memory experiment whereby each physical qubit is subject to a single-qubit depolarising error channel with a variable physical error rate $p \in [0.01,0.75)$. The goal is to compare the error suppression strengths between a logical qubit encoded in the repetition code, the virtual quantum repetition code, and the unrotated surface code. We consider a code-capacity error model, where stabiliser measurements are assumed noiseless. We study logical error rates defined by $p_{\mathrm{L}}(O) \vcentcolon = |1 - \langle O\rangle|/2$ for both $O \in \{X_{\mathrm{L}},Z_{\mathrm{L}}\}$ for varying code distances $d \in \{1,3,5,7\}$, where the logical qubit is initialised in the $+1$ eigenstate of $O$.

While the number of physical qubits is low enough to perform exact simulations (in QuEST~\cite{jones2019quest}) for the repetition code and the virtual quantum repetition code, we resort to Monte Carlo simulations (using Stim~\cite{gidney2021stim}) for the surface code when $d \geq 3$ (see \cref{tab:comparison} for a summary of quantum resource requirements). In the latter case, error bars correspond to hypotheses within a factor of 1000 of the maximum likelihood hypothesis.

For the bit-flip repetition code and the virtual quantum repetition code, we initialise the logical register to $\ket{0}_{\mathrm{L}}=\ket{0}^n$ ($\ket{+}_{\mathrm{L}}=(1/\sqrt{2})(\ket{0}^n + \ket{1}^n)$) when measuring logical bit-(phase-)flip errors. For the surface code, we initialise to $\ket{0}_{\mathrm{L}}$ ($\ket{+}_{\mathrm{L}}$) by applying one round of perfect stabiliser measurement on $\ket{0}^n$ ($\ket{+}^n$). Each physical data qubit, excluding the control qubit for H-VEC and for parity checks, is then subject to a single-qubit depolarising error defined by $\mathcal{D}_p[\,\cdot\,] = (1-p)I\cdot I + \frac{p}{3}\sum_{G\in \{X,Y,Z\}} G\cdot G$ as considered in \cref{sec:example}. The logical qubit of each code is then finally measured in the logical $Z$ or $X$ basis.

The figure shows that the virtual quantum repetition code is able to correct for both bit-flip and phase-flip errors, and is also able to reach lower logical error rates than the unrotated surface code and even the $Z$-basis case (which corrects bit-flip errors) of the repetition code.

\subsection{\cref{fig: threshold-theory}}
The setting of the numerical simulation corresponding to \cref{fig: threshold-theory} is the same as that of \cref{fig: threshold-cc}. The difference is that, in this case, we restrict ourselves to the $Z$-basis case where the bit-flip repetition code is able to correct errors, and show the correspondence between the simulated results and the leading-order analytical expressions given in \cref{sec:example}. Thus, solid lines of the figure are exactly the same as those in \cref{fig: threshold-cc}, and dashed lines correspond to \cref{eqn:p_l_rep_x}~(left), \cref{eqn:p_l_vec}~(middle), and \cref{eqn:p_l_sur}~(right).

\subsection{\cref{fig: sampling-cost}}
\cref{fig: sampling-cost} shows the sampling overhead $C_Y$ as a function of the physical error rate $p \in [0.01,0.5)$ of the virtual quantum repetition code, again with all physical qubits experiencing the same local depolarising error channel under a code-capacity error model. The plot compares between the leading-order analytical expression given in \cref{eqn:vec_example_overhead} and the numerical simulation computing $\expval{X \otimes I}$. While the sampling overhead is intractable for high physical error rates, we can observe that it is manageable in the high-fidelity regime of practical interest.

\subsection{\cref{fig: vepp_sim}}
Both subfigures consider a variable level of noise $p \in (0.00,0.75]$ acting upon otherwise ideal Bell states $\ket{\Psi^+}$ via a single-qubit depolarising channel $\mathcal{D}_p$ applied to either branch of the two-qubit state. While the plot on the right assumes that all input Bell states are noisy, the plot on the left keeps the Bell pair used to perform the $Z\otimes Z$ parity check ideal. The fidelity $F=\mathrm{Tr}(\ketbra{\Psi^+}\rho_{\mathrm{pur}})/\mathrm{Tr}(\rho_{\mathrm{pur}})$ is plotted as a function of $p$ in both cases.

For every implementation considered, the circuits are exactly simulated. For both single- and two-round EPPs, the fidelity is measured by computing the expectation value of $O_{\mathrm{Bell}}=(1/4)(I\otimes I+X\otimes X-Y\otimes Y+Z\otimes Z)$. For all virtual EPPs, this is done by taking the expectation values of $X\otimes X\otimes O_{\mathrm{Bell}}$ and $X\otimes X \otimes I$, and taking their ratio. Post-selection is simulated using normalised projection operators.

\bibliographystyle{unsrt}
\bibliography{ref}

\begin{thebibliography}{10}

\bibitem{calderbank1996good}
A.~R. Calderbank and Peter~W. Shor.
\newblock Good quantum error-correcting codes exist.
\newblock {\em Phys. Rev. A}, 54:1098--1105, Aug 1996.

\bibitem{steane1996multiple}
Andrew Steane.
\newblock Multiple-particle interference and quantum error correction.
\newblock {\em Proceedings of the Royal Society of London. Series A: Mathematical, Physical and Engineering Sciences}, 452(1954):2551--2577, 1996.

\bibitem{breuckmannQuantumLowdensityParitycheck2021}
Nikolas~P. Breuckmann and Jens~Niklas Eberhardt.
\newblock Quantum low-density parity-check codes.
\newblock {\em PRX Quantum}, 2(4):40101, October 2021.

\bibitem{tillichQuantumLDPCCodes2014}
Jean-Pierre Tillich and Gilles Z{\'e}mor.
\newblock Quantum {{LDPC}} codes with positive rate and minimum distance proportional to the square root of the blocklength.
\newblock {\em IEEE Transactions on Information Theory}, 60(2):1193--1202, February 2014.

\bibitem{breuckmannConstructionsNoiseThreshold2016}
Nikolas~P. Breuckmann and Barbara~M. Terhal.
\newblock Constructions and noise threshold of hyperbolic surface codes.
\newblock {\em IEEE Transactions on Information Theory}, 62(6):3731--3744, June 2016.

\bibitem{panteleevQuantumLDPCCodes2022}
Pavel Panteleev and Gleb Kalachev.
\newblock Quantum {{LDPC}} codes with almost linear minimum distance.
\newblock {\em IEEE Transactions on Information Theory}, 68(1):213--229, January 2022.

\bibitem{panteleevAsymptoticallyGoodQuantum2022}
Pavel Panteleev and Gleb Kalachev.
\newblock Asymptotically good quantum and locally testable classical {{LDPC}} codes.
\newblock In {\em Proceedings of the 54th {{Annual ACM SIGACT Symposium}} on {{Theory}} of {{Computing}}}, {{STOC}} 2022, pages 375--388, New York, NY, USA, June 2022. Association for Computing Machinery.

\bibitem{dinurLocallyTestableCodes2022}
Irit Dinur, Shai Evra, Ron Livne, Alexander Lubotzky, and Shahar Mozes.
\newblock Locally testable codes with constant rate, distance, and locality.
\newblock In {\em Proceedings of the 54th {{Annual ACM SIGACT Symposium}} on {{Theory}} of {{Computing}}}, {{STOC}} 2022, pages 357--374, New York, NY, USA, June 2022. Association for Computing Machinery.

\bibitem{liu2025virtual}
Zhenhuan Liu, Xingjian Zhang, Yue-Yang Fei, and Zhenyu Cai.
\newblock Virtual channel purification.
\newblock {\em PRX Quantum}, 6:020325, May 2025.

\bibitem{cai2023quantum}
Zhenyu Cai, Ryan Babbush, Simon~C. Benjamin, Suguru Endo, William~J. Huggins, Ying Li, Jarrod~R. McClean, and Thomas~E. O'Brien.
\newblock Quantum error mitigation.
\newblock {\em Rev. Mod. Phys.}, 95:045005, Dec 2023.

\bibitem{liu2025quantum}
Kecheng Liu and Zhenyu Cai.
\newblock Quantum error mitigation for sampling algorithms, 2025.

\bibitem{knill1997theory}
Emanuel Knill and Raymond Laflamme.
\newblock Theory of quantum error-correcting codes.
\newblock {\em Phys. Rev. A}, 55:900--911, Feb 1997.

\bibitem{koczor2021exponential}
B\'alint Koczor.
\newblock Exponential error suppression for near-term quantum devices.
\newblock {\em Phys. Rev. X}, 11:031057, Sep 2021.

\bibitem{huggins2021virtual}
William~J. Huggins, Sam McArdle, Thomas~E. O'Brien, Joonho Lee, Nicholas~C. Rubin, Sergio Boixo, K.~Birgitta Whaley, Ryan Babbush, and Jarrod~R. McClean.
\newblock Virtual distillation for quantum error mitigation.
\newblock {\em Phys. Rev. X}, 11:041036, Nov 2021.

\bibitem{caiPracticalFrameworkQuantum2021}
Zhenyu Cai.
\newblock A practical framework for quantum error mitigation.
\newblock {\em arXiv}, (arXiv:2110.05389 [quant-ph]), October 2021.

\bibitem{krishnaFaulttolerantGatesHypergraph2021}
Anirudh Krishna and David Poulin.
\newblock Fault-tolerant gates on hypergraph product codes.
\newblock {\em Physical Review X}, 11(1):11023, February 2021.

\bibitem{linErrormitigatedQuantumMetrology2025}
Xiaodie Lin and Haidong Yuan.
\newblock Error-mitigated quantum metrology via probabilistic virtual purification.
\newblock {\em arXiv}, (arXiv:2506.07618), June 2025.

\bibitem{haner2021distributed}
Thomas H\"{a}ner, Damian~S. Steiger, Torsten Hoefler, and Matthias Troyer.
\newblock Distributed quantum computing with {QMPI}.
\newblock In {\em Proceedings of the International Conference for High Performance Computing, Networking, Storage and Analysis}, SC '21, New York, NY, USA, 2021. Association for Computing Machinery.

\bibitem{itogawa2025efficient}
Tomohiro Itogawa, Yugo Takada, Yutaka Hirano, and Keisuke Fujii.
\newblock Efficient magic state distillation by zero-level distillation.
\newblock {\em PRX Quantum}, 6:020356, Jun 2025.

\bibitem{pino2021demonstration}
J.~M. Pino, J.~M. Dreiling, C.~Figgatt, J.~P. Gaebler, S.~A. Moses, M.~S. Allman, C.~H. Baldwin, M.~Foss-Feig, D.~Hayes, K.~Mayer, C.~Ryan-Anderson, and B.~Neyenhuis.
\newblock Demonstration of the trapped-ion quantum ccd computer architecture.
\newblock {\em Nature}, 592(7853):209--213, Apr 2021.

\bibitem{nickerson2013topological}
Naomi~H. Nickerson, Ying Li, and Simon~C. Benjamin.
\newblock Topological quantum computing with a very noisy network and local error rates approaching one percent.
\newblock {\em Nature Communications}, 4(1):1756, Apr 2013.

\bibitem{nickerson2014freely}
Naomi~H. Nickerson, Joseph~F. Fitzsimons, and Simon~C. Benjamin.
\newblock Freely scalable quantum technologies using cells of 5-to-50 qubits with very lossy and noisy photonic links.
\newblock {\em Phys. Rev. X}, 4:041041, Dec 2014.

\bibitem{horsman2012surface}
Dominic Horsman, Austin~G Fowler, Simon Devitt, and Rodney~Van Meter.
\newblock Surface code quantum computing by lattice surgery.
\newblock {\em New Journal of Physics}, 14(12):123011, dec 2012.

\bibitem{fowler2019low}
Austin~G. Fowler and Craig Gidney.
\newblock Low overhead quantum computation using lattice surgery, 2019.

\bibitem{litinski2019game}
Daniel Litinski.
\newblock A {G}ame of {S}urface {C}odes: {L}arge-{S}cale {Q}uantum {C}omputing with {L}attice {S}urgery.
\newblock {\em {Quantum}}, 3:128, March 2019.

\bibitem{vuillot2019code}
Christophe Vuillot, Lingling Lao, Ben Criger, Carmen García~Almudéver, Koen Bertels, and Barbara~M Terhal.
\newblock Code deformation and lattice surgery are gauge fixing.
\newblock {\em New Journal of Physics}, 21(3):033028, mar 2019.

\bibitem{jeon2026quantum}
Minjun Jeon and Zhenyu Cai.
\newblock Quantum error correction on error-mitigated physical qubits, 2026.

\bibitem{shalby2025optimized}
Mohamed~A. Shalby, Renyu Wang, Denis Sedov, and Leonid~P. Pryadko.
\newblock Optimized noise-resilient surface code teleportation interfaces.
\newblock {\em Phys. Rev. A}, 112:L020403, Aug 2025.

\bibitem{jacinto2026network}
Hugo Jacinto, \'Elie Gouzien, and Nicolas Sangouard.
\newblock Network requirements for distributed quantum computation.
\newblock {\em Phys. Rev. Res.}, 8:013205, Feb 2026.

\bibitem{haug2025lattice}
Trond~Hjerpekjøn Haug, Timo Hillmann, Anton~Frisk Kockum, and Raphaël~Van Laer.
\newblock Lattice surgery with {B}ell measurements: {M}odular fault-tolerant quantum computation at low entanglement cost, 2025.

\bibitem{huo2022dual}
Mingxia Huo and Ying Li.
\newblock Dual-state purification for practical quantum error mitigation.
\newblock {\em Phys. Rev. A}, 105:022427, Feb 2022.

\bibitem{jones2019quest}
Tyson Jones, Anna Brown, Ian Bush, and Simon~C. Benjamin.
\newblock {Q}u{EST} and high performance simulation of quantum computers.
\newblock {\em Scientific Reports}, 9(1):10736, Jul 2019.

\bibitem{jones2020questlink}
Tyson Jones and Simon Benjamin.
\newblock {Q}u{EST}link—{M}athematica embiggened by a hardware-optimised quantum emulator.
\newblock {\em Quantum Science and Technology}, 5(3):034012, may 2020.

\bibitem{gidney2021stim}
Craig Gidney.
\newblock Stim: a fast stabilizer circuit simulator.
\newblock {\em {Quantum}}, 5:497, July 2021.

\bibitem{kay2023tutorial}
Alastair Kay.
\newblock Tutorial on the {Q}uantikz package, 2023.

\bibitem{langer2005long}
C.~Langer, R.~Ozeri, J.~D. Jost, J.~Chiaverini, B.~DeMarco, A.~Ben-Kish, R.~B. Blakestad, J.~Britton, D.~B. Hume, W.~M. Itano, D.~Leibfried, R.~Reichle, T.~Rosenband, T.~Schaetz, P.~O. Schmidt, and D.~J. Wineland.
\newblock Long-lived qubit memory using atomic ions.
\newblock {\em Phys. Rev. Lett.}, 95:060502, Aug 2005.

\bibitem{sepiol2019probing}
M.~A. Sepiol, A.~C. Hughes, J.~E. Tarlton, D.~P. Nadlinger, T.~G. Ballance, C.~J. Ballance, T.~P. Harty, A.~M. Steane, J.~F. Goodwin, and D.~M. Lucas.
\newblock Probing qubit memory errors at the part-per-million level.
\newblock {\em Phys. Rev. Lett.}, 123:110503, Sep 2019.

\bibitem{seis2023balancing}
Yannick Seis, Benjamin~J. Brown, Anders~S. S\o{}rensen, and Joseph~F. Goodwin.
\newblock Improving trapped-ion-qubit memories via code-mediated error-channel balancing.
\newblock {\em Phys. Rev. A}, 107:052417, May 2023.

\bibitem{shi2025stabilizer}
Yu~Shi, Ashlesha Patil, and Saikat Guha.
\newblock Stabilizer entanglement distillation and efficient fault-tolerant encoders.
\newblock {\em PRX Quantum}, 6:010339, Mar 2025.

\bibitem{dur2007entanglement}
W~Dür and H~J Briegel.
\newblock Entanglement purification and quantum error correction.
\newblock {\em Reports on Progress in Physics}, 70(8):1381, jul 2007.

\bibitem{bennett1996purification}
Charles~H. Bennett, Gilles Brassard, Sandu Popescu, Benjamin Schumacher, John~A. Smolin, and William~K. Wootters.
\newblock Purification of noisy entanglement and faithful teleportation via noisy channels.
\newblock {\em Phys. Rev. Lett.}, 76:722--725, Jan 1996.

\bibitem{deutsch1996quantum}
David Deutsch, Artur Ekert, Richard Jozsa, Chiara Macchiavello, Sandu Popescu, and Anna Sanpera.
\newblock Quantum privacy amplification and the security of quantum cryptography over noisy channels.
\newblock {\em Phys. Rev. Lett.}, 77:2818--2821, Sep 1996.

\bibitem{yuan2024virtual}
Xiao Yuan, Bartosz Regula, Ryuji Takagi, and Mile Gu.
\newblock Virtual quantum resource distillation.
\newblock {\em Phys. Rev. Lett.}, 132:050203, Feb 2024.

\bibitem{yamamoto2024virtual}
Kaoru Yamamoto, Yuichiro Matsuzaki, Yasunari Suzuki, Yuuki Tokunaga, and Suguru Endo.
\newblock Virtual entanglement purification via noisy entanglement, 2024.

\bibitem{chamberland2020very}
Christopher Chamberland and Kyungjoo Noh.
\newblock Very low overhead fault-tolerant magic state preparation using redundant ancilla encoding and flag qubits.
\newblock {\em npj Quantum Information}, 6(1):91, Oct 2020.

\bibitem{gidney2024magic}
Craig Gidney, Noah Shutty, and Cody Jones.
\newblock Magic state cultivation: growing {T} states as cheap as {CNOT} gates, 2024.

\bibitem{tsubouchi2023virtual}
Kento Tsubouchi, Yasunari Suzuki, Yuuki Tokunaga, Nobuyuki Yoshioka, and Suguru Endo.
\newblock Virtual quantum error detection.
\newblock {\em Phys. Rev. A}, 108:042426, Oct 2023.

\bibitem{bennett1996mixed}
Charles~H. Bennett, David~P. DiVincenzo, John~A. Smolin, and William~K. Wootters.
\newblock Mixed-state entanglement and quantum error correction.
\newblock {\em Phys. Rev. A}, 54:3824--3851, Nov 1996.

\end{thebibliography}

\end{document}